\documentclass[12pt,a4paper,oneside]{article}
\usepackage{jheppub}
\usepackage{subfloat}
\usepackage{subfig}
\usepackage{amsmath,amsfonts,amssymb}
\usepackage{graphicx}
\usepackage{siunitx}
\usepackage{xspace}
\usepackage{braket}
\usepackage{mathtools}
\usepackage{hyperref}
\usepackage{comment}
\usepackage[normalem]{ulem}
\usepackage[capitalize]{cleveref}
\hypersetup{
    colorlinks=true,
    linkcolor=blue,
    citecolor=blue,
    urlcolor=blue
}

\binoppenalty=\maxdimen
\relpenalty=\maxdimen 

\graphicspath{{./graphics/}}

\def\bea{\begin{eqnarray}}
\def\eea{\end{eqnarray}}
\def\l{\left}
\def\r{\right}

\def\nn{\nonumber\\}

\def\bea{\begin{eqnarray}}
\def\eea{\end{eqnarray}}
\def\l{\left}
\def\r{\right}

\def\nn{\nonumber\\}
\def\bit{\begin{itemize}}
\def\eit{\end{itemize}}
\def\l{\left}
\def\r{\right}

\def\baa{\begin{array}}
\def\eaa{\end{array}}
\def\phys{{\rm phys} }


\title{ Bounds from $D/H$ on baryogenesis models
}
\preprint{SISSA 07/2026/FISI}

\author{Aleksandr Azatov}
\author{and Bruno Missoni}

\affiliation[]{SISSA, International School for Advanced Studies, Via Bonomea 265, 34136, Trieste, Italy}
\affiliation[]{INFN Sezione di Trieste, Via Bonomea 265, 34136, Trieste, Italy}
\affiliation[]{IFPU, Institute for Fundamental Physics of the Universe, Via Beirut 2, 34014 Trieste, Italy
}

\emailAdd{aleksandr.azatov@sissa.it}
\emailAdd{bruno.missoni@sissa.it}

\abstract{We review the constraints on baryon inhomogeneities derived from measurements of the deuterium abundance, $D/H$, and apply them to a range of baryogenesis models. In particular, we derive bounds on electroweak baryogenesis as well as on more exotic scenarios. Our results show that, across most of the relevant parameter space, electroweak baryogenesis remains largely unconstrained by current and foreseeable 
$D/H$ measurements. By contrast, the constraints on alternative scenarios are significantly stronger and can exclude regions of parameter space that would otherwise remain viable.
}
\begin{document}

 \maketitle\flushbottom
\newpage
\section{Introduction}
One of the fundamental problems in cosmology is understanding the origin of the observed matter-antimatter asymmetry in the Universe. This asymmetry is commonly parametrized by the baryon-to-entropy ratio, measured to be \cite{Planck:2015fie,Planck:2018vyg}
\begin{equation}
Y_B=\frac{n_B-n_{\bar B}}{s}\bigg|_0=(8.75\pm0.23)\times10^{-11},
\end{equation}
or, equivalently, by
$\eta\approx(n_{B,0}-n_{\bar{B},0})/n_{\gamma,0}\approx6.1\times10^{-10}$. 
This quantity can be determined independently from the angular power spectrum of the cosmic microwave background (CMB) and from the primordial abundances of light elements, once combined with the theory of Big Bang Nucleosynthesis (BBN) \cite{Yeh:2022heq}. The agreement between these two independent determinations is one of the major successes of modern cosmology.\newline
\newline
The BBN determination of the baryon asymmetry is primarily driven by measurements of the deuterium abundance, i.e. $D/H$. Recently, Ref. \cite{Bagherian:2025puf} pointed out that 
these measurements can also be used to constrain spatial inhomogeneities in the baryon abundance during the BBN epoch. The 
reason is that the dependence of the deuterium abundance on $\eta$ is non-linear (see for example \cite{Mukhanov:2005sc}), so fluctuations in the baryon abundance during 
BBN necessarily induce deviations in the predicted value of $D/H$. Therefore, precise 
measurements of $D/H$ can be used to place constraints on such fluctuations. 
Ref. \cite{Bagherian:2025puf} has proposed to use these constraints to test models of baryogenesis where the baryon production is not uniform in space. Interestingly, using  $D/H$ bounds, we can 
examine baryogenesis models that occurred during the earlier epoch. Indeed, all we need to check is 
whether the inhomogeneities in baryon asymmetry will be washed out by diffusion before the beginning of BBN. As emphasized in \cite{Bagherian:2025puf}, particles diffuse more efficiently at lower temperatures. In particular, the (comoving) diffusion lengths of protons and neutrons at the onset of BBN are \cite{Bagherian:2025puf}
\begin{equation}
    d_n\sim10^7\text{ cm}, \quad d_p\sim 10^4\text{ cm}.
\end{equation}
If these diffusion lengths are compared with the comoving Hubble radius at the electroweak scale ($T\sim 100 {\rm GeV}$),
\begin{equation}
L=(aH)^{-1}\sim 3\times 10^5 \text{ cm},
\end{equation}
it becomes clear that these observations may provide a way to test baryogenesis models in which the baryon asymmetry is generated at energies as high as ${\cal O}({\rm TeV})$. 
\newline\newline
In this paper, we have extended the results of Ref. \cite{Bagherian:2025puf} by presenting a detailed analysis of $D/H$ constraints on a few  explicit baryogenesis models. 
We start with the electroweak baryogenesis scenario (see for reviews \cite{Trodden:1998ym,Cline:2006ts,Riotto:1998bt,Riotto:1999yt,Dine:2003ax,White:2016nbo,Garbrecht:2018mrp,Bodeker:2020ghk,Barni:2025ifb,vandeVis:2025efm}).
In this case, bubbles of the broken phase percolate and fill the entire space. In these models, the generation of the baryon asymmetry of the Universe (BAU) is associated with a macroscopic length scale, set by the bubble size, which makes the scenario potentially interesting for $D/H$ bounds. However, we show that inhomogeneities in the baryon abundance arise only from the decrease in temperature during bubble expansion. As a result, the corresponding bounds turn out to be very weak in most of the parameter space.
Next, we 
proceed to more exotic BAU generation mechanisms in first-order phase transitions (FOPTs):
heavy particle production in  bubble-bubble collisions \cite{Cataldi:2024pgt,Cataldi:2025nac}, and heavy particle production in  plasma-bubble collisions \cite{Azatov:2021irb,Azatov:2022tii,Baldes:2021vyz}.
In this case, the bounds turn out to be more interesting due to geometry of the mechanism itself. Finally, we discuss baryogenesis models with domain walls \cite{Daido:2015gqa,Mariotti:2024eoh,Azzola:2024pzq,Schroder:2024gsi,Abel:1995uc,Brandenberger:1996st}. In this scenario, the bounds are particularly strong, since the length scale of inhomogeneities is comparable to the Hubble scale during the DW evolution.

The paper is organized as follows: in Section \ref{sec:review} we briefly review the bounds from $D/H$ and the limiting cases on length scales of inhomogeneities  compared to the diffusion lengths of protons and neutrons.
In Section \ref{Section: Electroweak baryogenesis},
\ref{Section: Production via bubble collisions}, \ref{Section: Production via relativistic bubble walls} and \ref{Section: Domain-wall baryogenesis} we present the constraints on four baryogenesis scenarios mentioned above. We summarize our results in Section \ref{Summary and Conclusions}.

\section{$D/H$ constraints on the baryon inhomogeneities}
\label{sec:review}
In this section, we briefly review the bounds on  the spatial inhomogeneities presented in \cite{Bagherian:2025puf}.  We will use exactly the same notation and conventions as the original reference and normalize the scale factor to be equal to 
\bea
a(T=1 \text{ MeV})=1.
\eea
The solution of the 
 diffusion equation gives us the following comoving diffusion lengths for protons and neutrons (see Section 2 of \cite{Bagherian:2025puf} or Appendix \ref{Appendix: Diffusion lengths for protons and neutrons})
\begin{equation}
    d_n\sim 10^7\text{ cm}, \quad d_p\sim 10^4\text{ cm}.
\end{equation}
Therefore, there could be three different cases depending on the ratios $L/d_{p,n}$:
\begin{itemize}
    \item \textbf{Case 1:} $L\gg d_n$, implying that both neutrons and protons do not homogenize before the start of BBN.
    \item \textbf{Case 2:} $ d_n \gg L \gg d_p$, protons homogenize before the start of BBN, but neutrons do not.
    \item \textbf{Case 3:} $d_p\gg L$, both protons and neutrons homogenize before the start of BBN.
\end{itemize}
Furthermore, the second case further divides into two subcases:
\begin{itemize}
    \item \textbf{Case (2a):} Neutron diffusion is not important during BBN, the deuterium production rate is large compared to diffusion, this happens for length scales around 6 km at the MeV temperature.
    \item  \textbf{Case (2b):} For inhomogeneities on scales which are larger than $d_p$, but for which neutron diffusion is important. 
\end{itemize}
While all the cases above have been treated extensively in \cite{Bagherian:2025puf}, here we provide a general summary of the results. The asymmetry in the comoving number density is parametrized as
\begin{equation}
    n_I(x)=n_I^0(1+\epsilon(x)),
\end{equation}
where $n_I^0$ is the homogeneous number density. The index $I$ is placed to distinguish different cases above (e.g. for Case 1: $n_1= n_B$ or Case 2: $n_2= n_p$, etc.). Running the BBN codes (e.g. PRIMAT \cite{Pitrou:2019nub} or PRyMordial \cite{Burns:2023sgx}) for different values of $\epsilon$ gives us the dependence of the $D/H$ ratio on $\epsilon$
\begin{equation}
    \frac{D}{H}\propto\frac{1}{(1+\epsilon)^a},
\end{equation}
where $a$ is some real number. At the same time, the measured quantity is
\bea
 \frac{\langle D\rangle}{\langle H\rangle } \simeq&&\frac{D|_{1+\epsilon}+D|_{1-\epsilon}}{H|_{1+\epsilon}+H|_{1-\epsilon}}=\frac{1}{2}\l[(1+\epsilon)\l.\frac{D}{H}\r|_{1+\epsilon}+(1-\epsilon)\l.\frac{D}{H}\r|_{1-\epsilon}\r]\nn 
\simeq && \l.\frac{D}{H}\r|_{\epsilon=0}+\epsilon^2 \l(\frac{D}{H}\r)'\bigg |_{\epsilon=0}
\eea
Thus, we can translate the bounds on $D/H$ to the bounds on the magnitude of the fluctuations. It turns out that for all the cases above, the bound on the RMS-fluctuation remains roughly the same 
\begin{equation}
    \epsilon_{RMS}\lesssim0.3,
\end{equation}
in order to stay within the most conservative error estimate of the $D/H$-ratio. 
\begin{equation}
    \frac{D}{H}=(2.53\pm0.1)\times10^{-5}.
\end{equation}
For reference, we will also report bounds assuming very optimistic errors on $D/H$ abundance determination\footnote{Such precision can be achieved by CMB 
measurements. However, current precision of nuclear reaction rates will increase the uncertainty to $0.01\to 0.03$ \cite{Yeh:2022heq}.}: 
$$(2.53\pm 0.01)\times10^{-5}$$
In this case, the bound on RMS fluctuations becomes roughly:
\bea
\epsilon_{RMS}^{optimistic} \lesssim 0.1.
\eea
Finally, let us illustrate the general strategy for evaluating $\epsilon_{RMS}$ here, and we leave   the details of the calculations to Appendix \ref{Appendix: Different distributions}. We first solve the diffusion equation in comoving coordinates
\begin{equation}
    \frac{\partial 
\rho}
{\partial t}=D\nabla^2\rho,
\end{equation}
with some initial density profile $\rho_0(x)=\rho(x,t_i)$. This is the inhomogeneous density profile left behind by the baryogenesis model under consideration, which will depend on the geometry and the comoving length scale of inhomogeneities. The solution is most easily obtained in momentum space\footnote{As shown in \cite{Bagherian:2025puf} and our Appendix \ref{Appendix: Diffusion lengths for protons and neutrons}, adopting a Gaussian profile $e^{-d^2k^2/2}$ for protons is not entirely correct. However, introducing an appropriate non-Gaussian profile does not change the bounds all that much.}
\begin{equation}
    \rho(\vec{x},t)=\int\frac{d^3k}{(2\pi)^3}e^{-d^2k^2/2}e^{i\vec{k}\cdot\vec{x}} \tilde{\rho}_0(\vec{k}),
\end{equation}
where $\tilde{\rho}_0(\vec{k})$ is the Fourier transform of the initial density profile. After this we can easily obtain the RMS fluctuation
\begin{equation}
    \epsilon_{RMS}^2=\frac{\langle\rho^2\rangle-\langle\rho\rangle^2}{\langle\rho\rangle^2},
\end{equation}
 and compare it with the experimental bound.
\section{Electroweak baryogenesis}\label{Section: Electroweak baryogenesis}
We start with electroweak baryogenesis,  one of the most popular and motivated 
scenarios for explaining the origin of the baryon asymmetry of the universe (BAU). In this case, baryogenesis occurs during the electroweak phase 
transition, which is assumed to be first-order. CP-violating interactions between the bubble wall and the hot plasma of particles generate asymmetries between left-handed and right-handed particles/antiparticles. This asymmetry is subsequently converted into the baryon 
asymmetry by electroweak sphaleron processes, which act only on the electroweak doublet fields. One 
of the main advantages of the scenario is its minimality since there is no need for additional sources of  baryon number 
violation except the ones already present in the SM. At the same time, new physics is still required  to make the electroweak (EW) phase transition first-order. In this scenario, there is a characteristic macroscopic length scale
$R_\phys$, the mean bubble size at the collision. In this paper, we will always add the subscript ``phys" when we talk about physical distances, whereas distances without a subscript are considered comoving.\newline
\newline
This length is related  to the nucleation parameters as follows \cite{Enqvist:1991xw}
\bea
 R_\phys\approx(8\pi)^{1/3}v_w\beta^{-1},~~~\beta =\frac{d S}{dt}=(T H)\frac{d S}{dT},
\eea
Here $S$ is the bounce-action exponent which controls the tunnelling rate for the bubble nucleation. 
Therefore, the maximum scale on which inhomogeneities can appear is $R_\phys$. The abundance of $D/H$ will be modified if these inhomogeneities 
are not smoothed out by proton diffusion. The effectiveness of 
diffusion will be controlled by the ratio
\bea
\frac{R_\phys}{a(T) d_{p}}\propto
v_w \l(\frac{H}{\beta}\r)\times \frac{(H(T) a(T))^{-1}}{d_{p}}\sim v_{w} \l(\frac{H}{\beta}\r)\times \frac{100 \text{ GeV}}{ T},
\eea
and inhomogeneities are erased if this ratio is much smaller than 1.  From the above estimates, we can see that electroweak baryogenesis 
models are close to being  
within the relevant
parameter range, since the quantity
 $\beta/H > \mathcal{O}(10)$
for successful baryogenesis models. 
However, this discussion is too naive. The locally-produced baryon asymmetry is controlled by the temperature of the plasma and the speed of the wall (which is also a function of the temperature). If these remain the same during the whole motion of the 
bubble, the production will be spatially 
homogeneous.
In the following, we estimate the typical dependence on these parameters. Thus, baryon inhomogeneities can arise only from temperature variations during the phase transition.
In particular, the temperature can decrease by an order one factor due to the Hubble expansion during the 
duration of the phase transition. So in order to estimate the baryon 
inhomogeneities, it becomes crucial to understand the dependence of baryon 
asymmetry on the temperature. The dependence on the temperature will 
come from the following sources, which in return are functions of the 
temperature: sphaleron rate, bubble wall velocity, chemical potential near the wall, bubble wall profile and the rates of the various reactions in plasma. To keep the calculation under control, we have split it into two independent parts:
\begin{itemize}
\item i) dependence of the baryon asymmetry on the temperature and velocity: $\eta(T)$
\item ii) bounds on the baryon number inhomogeneities assuming that
\bea
\label{eq:eta-dep}
\l|\frac{\eta(T_1)-\eta(T_2)}{\eta(T_1)+\eta(T_2)}\r|={\rm Max}\l| \frac{d\ln \eta}{d\ln T}\r| \times \l|\frac{T_1-T_2}{T_1+T_2}\r|
\eea
which is sufficient to derive optimistic bounds on the baryon asymmetry. 
\end{itemize}
The explicit calculation of the baryon asymmetry $\eta (T)$ for some benchmark scenarios  is discussed  in Appendix \ref{Appendix: Transport equations}. However, what is important, we generically find
\bea
{\rm Max}\l| \frac{d\ln \eta}{d\ln T}\r|\sim 1-5. 
\eea
In the next subsection, we will present the constraints on electroweak baryogenesis following Equation \ref{eq:eta-dep} and assuming that the maximal value for the logarithmic derivative of $\eta$ takes values $(1,10,100)$.

\subsection{Bounds on electroweak baryogenesis}
Let us suppose that the phase transition starts at $T_n$ (nucleation temperature) and finishes at $T_p$ (percolation temperature). Various regions in space will have then BAU in the range between $\eta(T_n)$ and $\eta(T_p)$.
We can parametrize such a distribution using the following ansatz:
\begin{equation}
    \rho(\vec{x},T=T_p)=\frac{1}{2}\left[1+\frac{\eta(T_p)-\eta(T_n)}{\eta(T_p)+\eta(T_n)}\cos\left(\frac{x\pi}{R}\right)\cos\left(\frac{y\pi}{R}\right)\cos\left(\frac{z\pi}{R}\right)\right],
\end{equation}
where $R$ is the (comoving) average bubble radius at the end of the phase transition $R\equiv R_\phys/a(T_p)\approx v_w(8\pi)^{1/3}/(a(T_p)\beta)$ and $\rho$ is normalized to be $\langle{\rho}\rangle=1/2$ under spatial averaging. The trigonometric form is chosen for calculational convenience, since it simplifies the solution of the diffusion equations.
In this ansatz, the volumes associated with the asymmetries $\eta(T_p)$ and $\eta(T_n)$ are the same, so that the resulting 
inhomogeneity is maximal. In reality, however, we expect most of the baryon asymmetry to be produced 
towards the end of the phase transition, when the bubbles occupy 
most of the volume. This implies that 
the actual constraints are expected to be even weaker.
In this case, the diffusion equation can be solved analytically and we obtain
\begin{equation}
    \rho(\vec{x},T)=\frac{1}{2}\left[1+e^{-3d^2\pi^2/(2R^2)}\frac{\eta(T_p)-\eta(T_n)}{\eta(T_p)+\eta(T_n)}\cos\left(\frac{x\pi}{R}\right)\cos\left(\frac{y\pi}{R}\right)\cos\left(\frac{z\pi}{R}\right)\right].
\end{equation}
With  the respective averages equal to
\begin{equation}
    \langle \rho \rangle=\frac{1}{2}, \quad \langle \rho^2 \rangle=\frac{1}{4}\left[1+\frac{1}{8}e^{-3d^2\pi^2/R^2}\left(\frac{\eta(T_p)-\eta(T_n)}{\eta(T_p)+\eta(T_n)}\right)^2\right].
\end{equation}
Modifying the trigonometric ansatz for the baryon asymmetry leads only to an 
$\mathcal{O}(1)$ change in the prefactor multiplying the exponential. Applying the deuterium RMS bound gives
\begin{equation}
    e^{-3d^2\pi^2/R^2}\left(\frac{\eta(T_p)-\eta(T_n)}{\eta(T_p)+\eta(T_n)}\right)^2\bigg\rvert_{\text{Max}} < 8 \epsilon_{RMS}^2\simeq 0.72 (\to 0.08),
\end{equation}
where $(\to 0.08)$ refers to the optimistic $\epsilon_{RMS}<0.1$ scenario. 
The temperature variation $\Delta T=T_n-T_p$ during the phase transition is related to its duration and as a consequence to the bubble radius $R$.
Since $R$ was defined in comoving coordinates, it must be equal to\footnote{The velocity $v_w$ is the physical one. Although it is generally temperature dependent here we treat it as a free parameter.} $R=v_w\Delta t/a(T_p)$. Using the time-temperature relation in the radiation-dominated era
\begin{equation}
    t=\frac{1}{2H(T)} \quad \rightarrow \quad \Delta t=-\frac{\Delta T}{H(T_{p})T_{p}} \quad \rightarrow \quad R\approx -\frac{v_w\Delta T}{H(T_p)T_pa(T_p)},
\end{equation}
we then obtain
\begin{equation}
    \frac{3d^2\pi^2}{R^2}\approx \frac{3d^2\pi^2H^2(T_p)T_p^2a^2(T_p)}{v_w^2\Delta T^2}.
\end{equation}
Then we can use the relation for maximal baryon inhomogeneities from Eq. \eqref{eq:eta-dep} to get
\begin{equation}
\label{eq:dTTMax}
 \left(
 \frac{\Delta T}{2 T}\times {\rm Max}\ \frac{d\ln \eta}{d\ln T}
 \right)^2  \exp\left[-\frac{3d^2\pi^2H^2(T_p)T_p^2a^2(T_p)}{v_w^2\Delta T^2}\right]\lesssim 0.72\ (\to 0.08).
\end{equation}
\begin{figure}[t!]
    \centering
    \subfloat[$\epsilon_{RMS}<0.3$]{
        \includegraphics[width=0.4\linewidth]{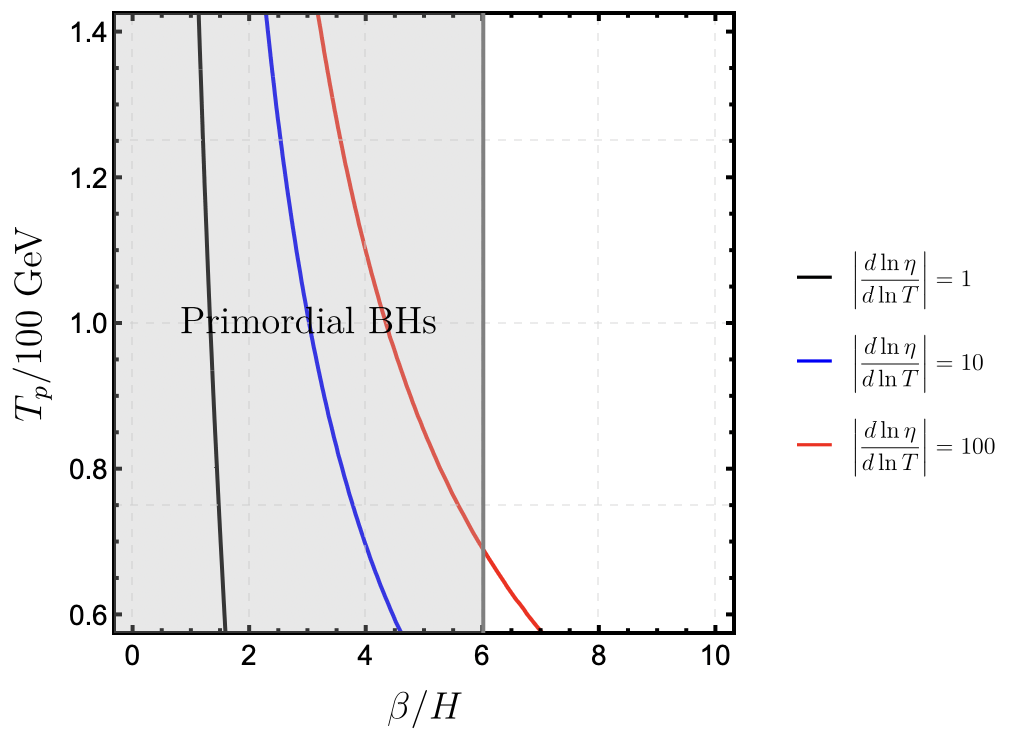}
        \label{fig:alphacont1}
    }
    \hfill
    \subfloat[$\epsilon_{RMS}<0.1$]{
        \includegraphics[width=0.4\linewidth]{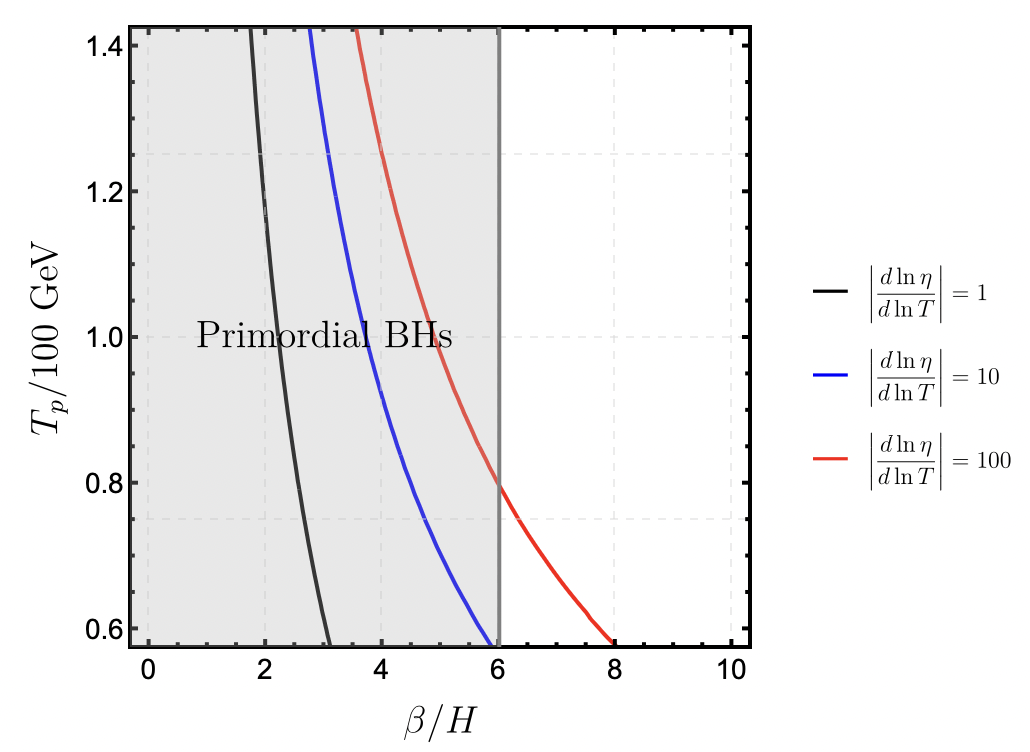}
        \label{fig:alphacont2}
        }
    \caption{Contours saturating the bound \eqref{eq:maineq-EW} for different values of the maximal derivative $\text{Max }|d\ln\eta/d\ln T|$, where $v_w=0.1$. The allowed parameter space is placed to the right of the curves and the grey area is excluded by primordial BH production \cite{Gouttenoire:2023naa,Liu:2021svg}.}
    \label{fig:combined_EWBG2}
\end{figure}\newline
Now we can relate $\Delta T$ to the bubble nucleation parameter $\beta$
\begin{equation}
R_{phys}\approx(8\pi)^{1/3}v_w\beta^{-1}\approx -\frac{v_w\Delta T}{H(T_p)T_p}\quad \rightarrow\quad \Delta T=-\frac{(8\pi)^{1/3}H(T_p)T_p}{\beta}.
\end{equation}
Plugging this expression in the Eq. \ref{eq:dTTMax} we obtain the following bound
\begin{equation}
    \left(\frac{H(T_p)}{\beta}\right)^2
    \left({0.5\times\rm Max}\frac{d\ln \eta}{d\ln T}\right)^2\exp\left[-3.45\frac{(d\ \beta \ a(T_p))^2}{v_w^2}\right]\lesssim 0.08 (\to 0.01).
\end{equation}
Setting $d=d_p$, where $d_p$ is the proton diffusion length we obtain: 
\begin{equation}
\label{eq:maineq-EW}
\text{Max} \left\lvert\frac{d\ln\eta(T)}{d\ln T}\right\rvert_{T=T_p}\lesssim 0.56\  (\to 0.19)\frac{\beta}{H(T_p)}\exp\left[\frac{2\times 10^{-3}}{v_w^2}\left(\frac{T_p}{100\text{ GeV}}\right)^2\left(\frac{\beta}{H(T_p)}\right)^2\right].
\end{equation}
The resulting bounds are shown in Figure
\ref{fig:combined_EWBG2}, where we have 
plotted the exclusion lines for the  
values of\footnote{Note that in realistic models one typically expects   ${\rm Max}\l| \frac{d\ln \eta}{d\ln T}\r|\in[1,5] $.} ${\rm Max}\l| \frac{d\ln \eta}{d\ln T}\r|=(1,10,100)$, for conservative and optimistic constraints 
on $\epsilon_{RMS}$. As expected from Eq. \eqref{eq:maineq-EW}, the bounds become stronger for relatively small values of $\beta/H$. In this regard, it is important to emphasize the primordial black hole production bound \cite{Gouttenoire:2023naa,Liu:2021svg},
 which excludes phase transitions with  $\beta/H\lesssim 6-8$.  We therefore conclude that  models with electroweak baryogenesis  remain largely 
unconstrained by BBN, unless the baryon 
asymmetry has an unrealistically strong 
temperature dependence and the 
percolation temperature is unusually low.\newline
\newline
In the following sections, we consider more exotic models for the generation of the BAU, for which the BBN constraints become more important.


\section{Production via bubble collisions}\label{Section: Production via bubble collisions}
In this section, we consider a different class of baryogenesis models, namely those in which the BAU is generated through heavy-particle production during bubble collisions \cite{Falkowski:2012fb,Katz:2016adq,Shakya:2023kjf,Mansour:2023fwj,Cataldi:2024pgt,Cataldi:2025nac}. In these kinds of models, bubbles accelerate to relativistic velocities, 
so that the characteristic energy at the moment of collision is much larger than the scale of the phase transition. As a result, heavy states with masses much larger than the critical temperature can be produced.
If this process involves CP violation and B-violating interactions, a successful baryogenesis scenario can be realized. 
\begin{figure}[t!]
\centering
\includegraphics[scale=0.6]{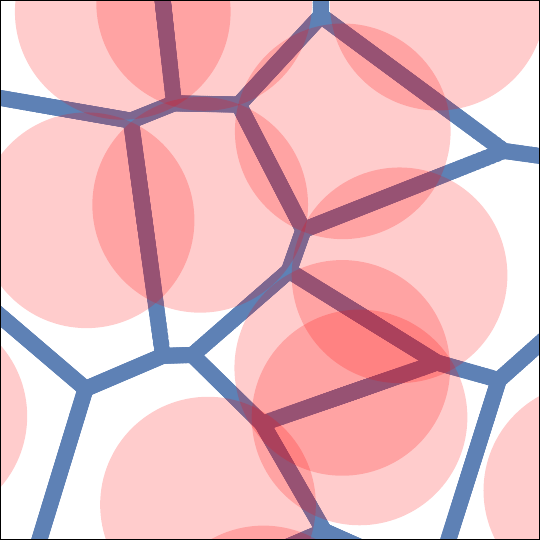}
\caption{Blue lines indicate the regions where the bubble collisions  take place and where the initial baryon asymmetry will be localized. For the illustration we have chosen simultaneous nucleation of the bubbles in the 2D plane.
\label{fig:bub-col}}
\end{figure}
An interesting feature relevant to the $D/H$ bounds is that the production of baryon asymmetry occurs only in the regions of space where bubbles have collided (see Fig. \ref{fig:bub-col}). Therefore, the baryon asymmetry is confined to very small regions, with the typical distance between these regions set by the inter-bubble distance $R_{\phys}$.

\subsection{Constraints on the scale of the phase transition}
Let us now proceed with the analysis of the $D/H$ bounds. Since particle production is strongly localized in space, we can model the baryon asymmetry profile at the end of the phase transition as a sum over $\delta$-functions\footnote{Note that this particular example was considered as a toy model in the Appendix of \cite{Bagherian:2025puf}. However, they did not push the analysis further to place the bound on FOPT parameters.}
\bea    \rho(\vec{r},t=0)=\frac{1}{N^3}\sum_{n_x,n_y,n_z=-N/2}^{N/2}\delta^3(\vec{r}-R(n_x\hat{x}+n_y\hat{y}+n_z\hat{z})),
\eea
where $R\equiv R_\phys/a(T_p)$ is the average comoving distance between two bubbles. The solution to the diffusion equation with this initial condition can be found in Appendix \ref{Appendix: delta function inhomogeneities}. Here we just report the RMS fluctuation
\begin{equation}\label{delta func bound}
    \epsilon_{RMS}^2=\left(\frac{R}{2\sqrt{\pi} d}\sum_{p=-\infty}^{\infty}e^{-R^2p^2/4d^2}\right)^3-1.
\end{equation}
From Figure \ref{delta func bound}, it can be seen that the deuterium bound is satisfied when
\begin{equation}
    R\lesssim 3 \ (\to2.44)d,
\end{equation}
where $(\to2.44)$ refers to the optimistic $\epsilon_{RMS}<0.1$ scenario.
Converting to the physical length scale
\begin{equation}
   R_\phys < 3(\to2.44)\ a(T_{PT}) d(T_{BBN})\approx 3 \ (\to2.44) a(T_{PT}) \ 10^4 \text{cm}
\end{equation}
where $T_{PT}$ is some temperature at the scale of the phase transition and we used the diffusion length of protons $\sim 10^4$ cm as a more stringent bound. Using the same normalization as in \cite{Bagherian:2025puf} ($a(T_0)=1, T_0=1$ MeV) we have that
\begin{equation}
    a(T)=\frac{T_0}{T}\left(\frac{g_{\ast S}(T_0)}{g_{\ast S}(T)}\right)^{1/3}
\end{equation}
and we get
\begin{equation}
\begin{split}
    R_\phys&<\frac{7.11 \ (\to5.78)\times10^{14}}{T_{PT}}.
\end{split}
\end{equation}
\begin{figure}
    \centering
    \includegraphics[width=0.5\linewidth]{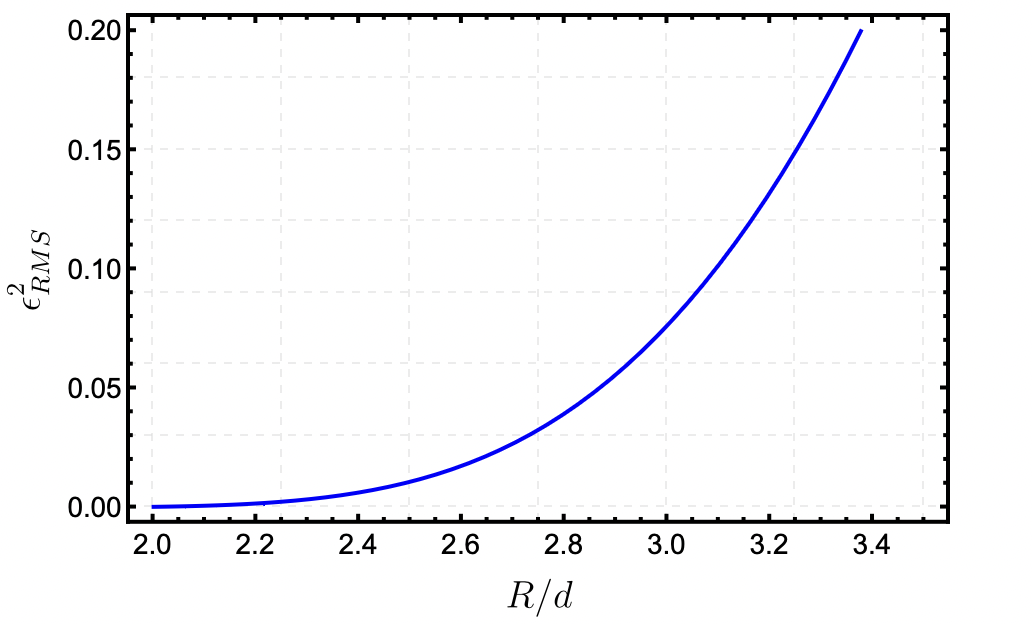}
    \caption{$\epsilon_{RMS}^2$ for $\delta$-function inhomogeneities as in \eqref{delta func bound}.}
    \label{fig:delta func bound}
\end{figure}\newline
We multiply by $H(T_{PT})\sim \frac{1}{\sqrt{3}M_{pl}}\langle \phi\rangle^2$, where $\langle \phi \rangle$ is the VEV of the scalar driving the FOPT and $M_{Pl}=2.4\cdot10^{18}\text{ GeV}$ is the reduced Planck mass. We assume strong supercooling since such FOPTs generally lead to relativistic speeds. Then we get
\begin{equation}
    \frac{7.11\ (\to5.78)\cdot10^{14}}{T_{PT}}H(T_{PT})\sim \frac{7.11\ (\to5.78)\cdot10^{14}}{T_{PT}}\frac{\langle \phi\rangle^2}{\sqrt{3}M_{Pl}}\approx 0.017 \ (\to0.014)\frac{\langle\phi\rangle}{100 \text{ GeV}} \frac{\langle\phi\rangle}{T_{PT}}.
\end{equation}
The bound is then given by 
\begin{equation}
    R_\phys H(T_{PT})\lesssim 0.017 \ (\to0.014)\frac{\langle\phi\rangle}{100 \text{ GeV}} \frac{\langle\phi\rangle}{T_{PT}}.
\end{equation}
Using the fact that $R_{\rm phys}\approx (8\pi)^{1/3}\beta^{-1}$, we get
\begin{equation}
\label{eq:results-bb}
    \frac{\beta}{H(T_{PT})}\gtrsim 172.3\ (\to209.2) \frac{100\text{ GeV}}{\langle\phi\rangle}\frac{T_{PT}}{\langle \phi \rangle}.
\end{equation}
We can see that models with the phase transition scale (critical temperature) below 100 GeV become tightly constrained.

\section{Production via relativistic bubble walls}\label{Section: Production via relativistic bubble walls}
In the case of relativistic bubbles, another interesting baryogenesis model has been proposed in \cite{Azatov:2020ufh,Azatov:2021irb,Baldes:2021vyz}. Similarly to the 
previous section, the baryogenesis proceeds via heavy-particle production, but in this case it originates from 
the bubble collisions with the surrounding plasma particles. The 
probability for the production of heavy states roughly scales as 
\begin{equation}
    \mathcal{P}_{ \text{heavy}}\sim\frac{\langle\phi\rangle^2}{M_{\ast}^2}\theta(\gamma_wT_n\langle\phi\rangle-M_{\ast}^2),
\end{equation}
where $\theta$ is the Heaviside theta function, $T_n$ is the nucleation temperature and $M_{\ast}$ is the mass of the heavy state.
\begin{figure}[t!]
    \centering
    \includegraphics[width=0.35\linewidth]{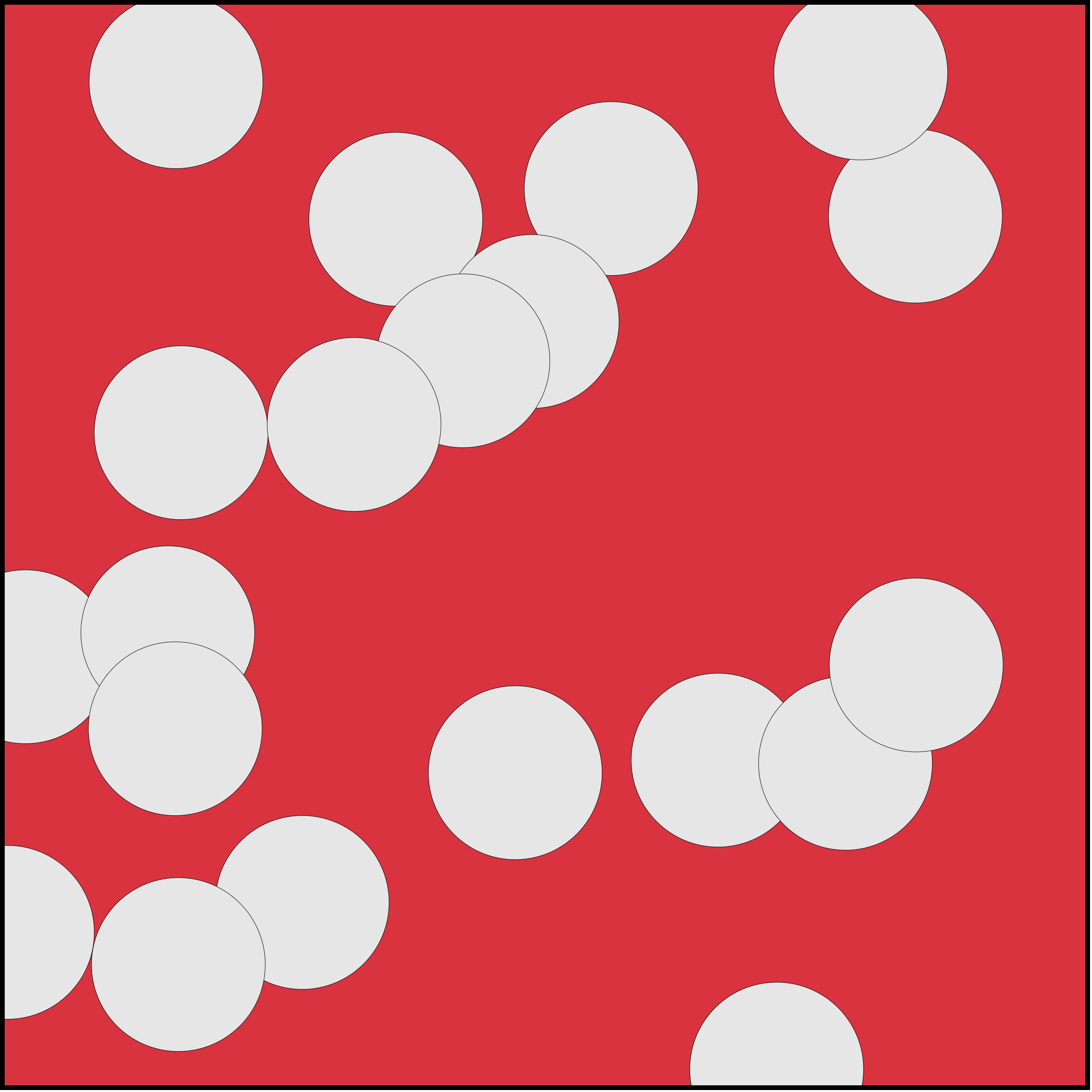}
    \caption{Illustration of the type of inhomogeneities left by the relativistic bubble walls. We have represented the area where the BAU is non-zero in red.}
    \label{fig:empty holes}
\end{figure}
The appearance of the step function can be understood intuitively since the energy scale of the bubble plasma scales as $E\sim \sqrt{\gamma_wT_n\langle\phi\rangle}$, and it obviously must be larger than the masses of the particles produced. At the same time, this means that particle production  does not start until the bubble has achieved a sufficient Lorentz boost factor
\bea
\gamma_{\rm min}\sim \frac{M_*^2}{T_n \langle \phi \rangle}.
\eea
During the initial stages, the bubble expansion is accelerated and the bubble radius is proportional to its boost factor
\bea
\gamma\simeq \frac{2}{3}\frac{R_{\rm phys}}{R^0_{ \rm phys}},
\eea
where $R^0_\phys$ is the bubble radius at the moment of nucleation and is typically of the order of $\langle \phi \rangle^{-1}$. We can see that particle production starts only once the radius of the bubble becomes larger than 
\bea
\label{eq:rmin}
R^{\rm min}_\phys\sim \frac{3}{2}\gamma_{\rm min} R^0_{\rm phys} \sim \frac{3}{2}\frac{M_*^2}{T_{PT}\langle\phi\rangle}R^0_{\rm phys}.
\eea
Thus, space will contain ``holes" of radius $R^{\min}_\phys$ with zero baryon asymmetry (see Figure \ref{fig:empty holes}). In the following, we will use the 
 deuterium bound to place restrictions on these types of models.
\subsection{$D/H$ bounds}
\begin{figure}[t!]
    \centering
    \includegraphics[width=0.5\linewidth]{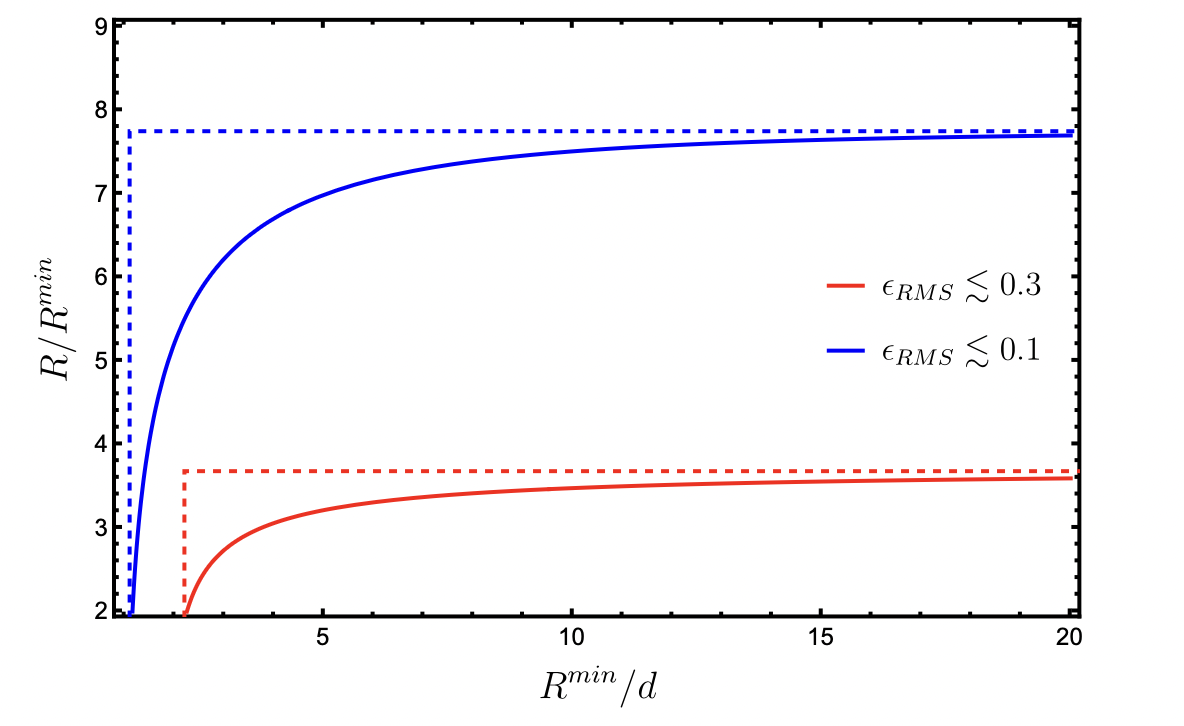}
    \caption{The curves representing the deuterium bound for relativistic bubble production with various precisions of $D/H$ measurements. The allowed region in the parameter space is placed above the curve. The dashed rectangles represent the approximations in \eqref{Rectangle approximation}.}
    \label{fig:allowed rel bubbles}
\end{figure}

We have argued that the baryon asymmetry distribution in space at the instance of the phase transition will 
be uniform apart from the empty ``holes" with radii  $\sim R^{\rm min}_\phys$. The typical distance between 
these ``holes" will be of the same order as the typical bubble size at the moment of collision. 
We will model this distribution by a lattice of empty spheres  with radius $R^{\rm min}_{\phys}$ and inter-sphere distance $R_\phys$. 
In comoving coordinates, introducing $R^{\rm min}\equiv R^{\rm min}_{\phys}/a$,  we get\footnote{Note that $R>2R^{\rm min}$ so that we are not in the bubble collision scenario.}
\bea 
\rho(r,t=0)=1-\sum_{\vec{n}} \theta(R^{\min}-\lvert \vec{x}-\vec{n}R\rvert),
\eea 
In this system we look at the bound in the parameter space of two dimensionless ratios: $R/R^{\rm min}$ and $R^{\rm min}/d$. The allowed parameter space is shown in Figure \ref{fig:allowed rel bubbles}.
The derivation of the curve is relegated to Appendix \ref{spheres}, however some main features can be understood using the following intuitive arguments:
\bit
\item $R^{\min}/d \gg  1$:  In this case 
diffusion effects cannot smooth out the baryon 
    asymmetry, and the constraint from $D/H$ becomes a requirement that 
    the total volume occupied by the empty ``holes" is sufficiently small. Numerically, this condition becomes an inequality:
\begin{equation}
    \frac{R}{R^{\rm min}}>3.68\ (\to7.75) .
\end{equation}
 Using Eq. \eqref{eq:rmin} and the relation  $R_{\rm phys}=(8\pi)^{1/3}\beta^{-1}$ (we take $v_w\approx1$).\begin{equation}
    \beta^{-1}\gtrsim1.86 \ (\to3.72)\frac{M_\ast^2}{T_{PT}\langle\phi\rangle}R^{ 0}_{\rm phys}.
\end{equation}
Inverting the inequality and dividing by $H\sim \frac{\langle\phi\rangle^2}{\sqrt{3}M_{Pl}}$ (assuming strong supercooling again), we obtain the following result
\begin{equation}
\begin{split}\label{bounds on betaover H}
    \frac{\beta}{H}&\lesssim 0.93\  (\to0.47)\frac{T_{PT}}{\langle \phi \rangle} \frac{M_{Pl}}{R^{0}_{\rm phys}M_\ast^2}\sim (\to0.5)\frac{T_{PT} M_{Pl}}{M_*^2},
\end{split}
\end{equation}
where in the last step we have assumed 
$R^0_{\phys}\sim \langle \phi \rangle^{-1}$. As expected, the bound is looser for smaller values of $M_*$, since in this limit the particle production starts earlier and the ``hole" size is smaller.
\item For small values of $R^{\rm min}/d \sim O(1)$ and for all values of $R$ such that $R> 2 R_{\min}$, the models satisfying\footnote{Note that it is important to stress the temperature at which we evaluate the diffusion length.}
\begin{equation}
    \frac{R^{\rm min}}{d(T_{BBN})}<2.2 \ (\to1.2).
\end{equation}
will pass the $D/H$ constraints. This implies that the diffusion is efficient enough to ``fill the empty holes". Then 
\begin{equation}
    R^{\rm min}=\frac{R^{\rm min}_{\rm phys}}{a(T_{PT})}\sim \frac{3}{2a(T_{PT})}\frac{M_\ast^2R^{0}_{\rm phys}}{T_{PT}\langle\phi\rangle},
\end{equation}
which leads to
\begin{equation}
    \frac{M_\ast^2R^{0}_{\rm phys}}{T_{PT}\langle\phi\rangle }\lesssim 1.47\ (\to0.8)\ a(T_{PT})  d_p( T_{BBN}).
\end{equation}
Using the proton diffusion length as a more stringent bound, we find
\begin{equation}\label{bounds on masses}
    \frac{M_\ast^2R^{0}_{\rm phys}}{\langle\phi\rangle}\lesssim3.47 \ (\to1.89) \times 10^{14}. 
\end{equation}
\eit
To summarize, in the limiting cases, we obtain two bounds: \eqref{bounds on betaover H} and \eqref{bounds on masses}, one of which must be satisfied.
Thus, the models satisfying the constraint
\bea\label{Rectangle approximation}
   M_*^2 \lesssim \text{ Max}\left[3.47\ (\to 1.89)\cdot 10^{14}\frac{\langle\phi\rangle}{R^{0}_{\rm phys}}, 0.93\ (\to 0.47)\frac{T_{PT}}{\langle\phi\rangle}\frac{M_{Pl}}{R^{0}_{\rm phys}} \times \left(\frac{H}{\beta}\right)\right].   \eea 
will pass the $D/H$ bounds. This constraint can be thought of as approximating the curve in Figure \ref{fig:allowed rel bubbles} by a rectangle (dashed lines in Figure \ref{fig:allowed rel bubbles}). 
\newline
\newline
Finally, we can write down a sufficient but not necessary condition:
\bea
\frac{R^{\rm }}{d} \lesssim 2.2.
\eea
In this case, no matter what the radius of the empty sphere is, diffusion will make the universe homogeneous. This condition becomes similar to the bounds obtained in the previous section (see Eq. \eqref{eq:results-bb})
\bea
\frac{\beta}{H}\gtrsim 100 \frac{T_{PT}}{\langle\phi\rangle}\frac{100 \text{GeV}} {\langle\phi\rangle}\l(\frac{10}{g_*}\r)^{1/2},
\eea
and we immediately see that the interesting region are the transitions below the 100 GeV scale.\newline
\newline
Finally, let us comment on the scenario in which the phase transition under discussion is the electroweak one (e.g. see Ref.~\cite{Azatov:2022tii}). In this case, 
due to the friction from vector boson emissions, the bubbles reach terminal velocity for relatively small values of 
the Lorentz boost factor $\gamma$. As a consequence, the radius of the  empty ``holes" always  satisfies
$R_\phys^{\min} \ll R_{\phys}$ and these models automatically satisfy the BBN bounds.

\section{Domain-wall baryogenesis}\label{Section: Domain-wall baryogenesis}
The final scenario considered in this paper is a class of BAU-generation models in which baryogenesis is driven by the evolution of the domain-wall network.
Domain walls (DWs) are topological defects that can arise when discrete symmetries are spontaneously broken. 
The space of vacua after the symmetry breaking is disconnected, and the domain 
walls interpolate between the different vacua of the theory. 
We will restrict ourselves to the simplest case where the broken discrete symmetry is $\mathbb{Z}_2$. 
The study of domain-wall evolution in the universe is usually performed numerically \cite{Press:1989yh,Kawano:1989mw,Avelino:2005kn,Leite:2011sc} (for more recent references, see \cite{Dankovsky:2024zvs,Blasi:2025tmn}).
The general picture presented in the literature is broadly the same: domains corresponding to different vacua form on small length scales and subsequently evolve toward the so-called scaling regime \cite{Kibble:1976sj}, in which there is approximately one domain wall per Hubble volume.
In terms of the comoving distances the scaling solution can be parametrized as follows
\bea
\xi_c=\epsilon  \eta, ~~\epsilon\simeq 0.5,
\eea 
where $\eta$ is the conformal time and $\xi_c$
is the comoving size characterizing the wall separation. Typical velocities of the domain walls  are found to be  
\bea
\gamma v\simeq 0.4.
\eea
The isotropy and homogeneity of our current universe suggest that the domain walls should have been annihilated by today \cite{Zeldovich:1974uw, Vilenkin:2000jqa}. Moreover, since the domain walls redshift 
slower than matter and radiation, they should disappear before they started dominating the universe energy budget not to 
spoil the expansion history of the universe and various measurable quantities (e.g. BBN abundances \cite{Azzola:2024pzq}). To 
resolve this, a small bias term in the potential $\delta V$ is introduced that favours one of the two degenerate vacua. 
When the bias term wins over the wall tension, domain walls annihilate and the universe remains in a single vacuum.\newline
\newline
Domain walls can also be used for baryogenesis and various models have been proposed \cite{Daido:2015gqa,Mariotti:2024eoh,Azzola:2024pzq,Schroder:2024gsi,Abel:1995uc,Brandenberger:1996st}. 
For the purpose of imposing the deuterium bound,
the key difference lies in the CP-violating term in the Lagrangian.
More precisely, it is determined by the power of the symmetry-breaking scalar $S$ entering the CP-violating interaction, namely:
\begin{itemize}
    \item \textbf{$\mathbb{Z}_2$ odd}: In this case, the CP violation and consequently the sign of the baryon asymmetry generated 
    will depend on the sign of $\langle S\rangle$ in the domain. During the 
    scaling solution, the domain wall 
    passes multiple times through a given point in comoving coordinates, generating baryon asymmetry of the opposite sign with every passage. Then on average no 
    baryon asymmetry will be generated 
    except for the last passage, which 
    happens at the instance of domain wall 
    annihilation. This discussion is very rough since there is a hidden assumption that the baryon production is independent of the temperature, otherwise contributions of different sign will not cancel explicitly.
    
    \item \textbf{$\mathbb{Z}_2$ even}: Here, the sign of the baryon asymmetry generated does not depend on the sign of 
    $\langle S\rangle$ in the domain, and the total baryon asymmetry must be added up.
\end{itemize}
While the bounds we obtain generally fall into the two cases mentioned above, we focus on models incorporating EW baryogenesis with domain walls \cite{Schroder:2024gsi,Azzola:2024pzq}, keeping the discussion as general as possible to allow straightforward application to other models. We will assume that the DWs have 
formed and reached the scaling regime at some temperature $T_{dw}$. After the electroweak phase transition, we can have 
electroweak symmetry restoration within the DWs, in \cite{Azzola:2024pzq} referred to as EW cores. Within the EW cores, the EW sphalerons are 
active and we can have baryon production. Due to the non-trivial DW profile and the CP-violating term, we have a 
mechanism similar to the one adopted in Section \ref{Section: Electroweak baryogenesis}. The chiral bias is created within 
the EW cores, where active sphalerons convert it to baryon asymmetry. 

We will show that the 
$D/H$
bounds in this class of models are very stringent (as noted in 
Ref.~\cite{Bagherian:2025puf}), since the characteristic scale of the
inhomogeneities is set by the distance 
between DWs, which is comparable to the Hubble scale itself.

\subsection{Production from the DW annihilation only}\label{Section: DW annihilation}
Consider the annihilation of DWs, which can occur if a small bias term is added to the $S$ potential, explicitly breaking the $\mathbb{Z}_2$ symmetry. At some point DWs start to annihilate and we suppose that the baryon asymmetry is generated only during the period of domain-wall annihilation. This is approximately the case if the 
CP-violation term depends on the sign of $\langle S \rangle$. During this annihilation, the DWs will sweep through the false vacuum regions and the baryon asymmetry generation will occur only in the domains with the false vacuum.
The distance between the DWs will be fixed by the scaling solution at the instance when the annihilation starts.\newline
\newline
This will leave an inhomogeneous distribution of baryons, which we model with a ``2D chess board" density profile\footnote{The reason that this is a 2D chess board is that domain walls in three dimensions typically have the shape of infinite sheets with varying curvature radius, as confirmed by numerical simulations \cite{Blasi:2025tmn,Kawano:1989mw,Leite:2011sc,Press:1989yh, Dankovsky:2024zvs}. Domain walls which form closed surfaces would collapse under their own surface tension, as there are no outward forces due to the two vacua being degenerate.}
\begin{equation}
    \rho(\vec{x},t)=\frac{1}{2}\left[1+\text{sgn}\left(\cos\left(\frac{\pi x}{D}\right)\cos\left(\frac{\pi y}{D}\right)\right)\right].
\end{equation}
The difference between this model and the other ones lies in the fact that $D$ is determined by the comoving horizon size at the annihilation temperature $T_{ann}$. The solution of the diffusion equation follows the usual scheme
described in Section \ref{sec:review}
and we leave the details of the calculations to Appendix \ref{Appendix: DW annhiliation}. Here we  report only the expression for the RMS fluctuation
\begin{equation}
    \epsilon_{RMS}^2=\left(\frac{8}{\pi^2}\sum_{m=0}^\infty \frac{1}{(2m+1)^2}e^{-d^2(2m+1)^2\pi^2/D^2}\right)^2.
\end{equation}
Imposing the deuterium bounds, we get
\begin{equation}
    \frac{d(T_{BBN})}{D}>0.317.
\end{equation}
Since the domain walls annihilate after reaching the scaling regime, $D$ is given by the comoving horizon size $D\sim 1/(aH)$ at the temperature of domain wall annihilation
\begin{equation}
    D\sim\frac{1}{(aH)}\bigg\rvert_{T=T_{ann}}=\frac{3}{\pi}\left(\frac{10}{g_*(T)}\right)^{1/2}\left(\frac{g_{\ast S}(T)}{g_{\ast S}(T_0)}\right)^{1/3}\frac{M_{Pl}}{T_0T_{ann}}.
\end{equation}
This leads to a bound on the temperature of domain-wall annihilation (using the diffusion length of protons)
\begin{equation}\label{annihilation temperature bound}
    T_{ann}\gtrsim1.7\text{ TeV}.
\end{equation}
Again, improving the $D/H$ measurements will not change the bounds all that much and taking $\epsilon_{RMS}<0.1$ we obtain
\begin{equation}
    T_{ann}\gtrsim2.5 \text{ TeV}.
\end{equation}
Note that this discussion was completely model independent and the only assumption was that BAU generation occurs only at the instance of annihilation.  If we apply it to the model in \cite{Azzola:2024pzq} where the baryon production should occur \textbf{after} the EW phase transition ($T_{EW} \sim 130$ GeV), we see that this is incompatible with the bound in \eqref{annihilation temperature bound}. Thus, we can see that electroweak baryogenesis with domain walls, where the CPV term is odd in the scalar field, possibly does not satisfy the bounds coming from $D/H$ measurements.

\subsection{Production from the scaling regime}\label{Section: DW scaling regime}

\begin{figure}[t!]
    \centering
    \includegraphics[width=0.5\linewidth]{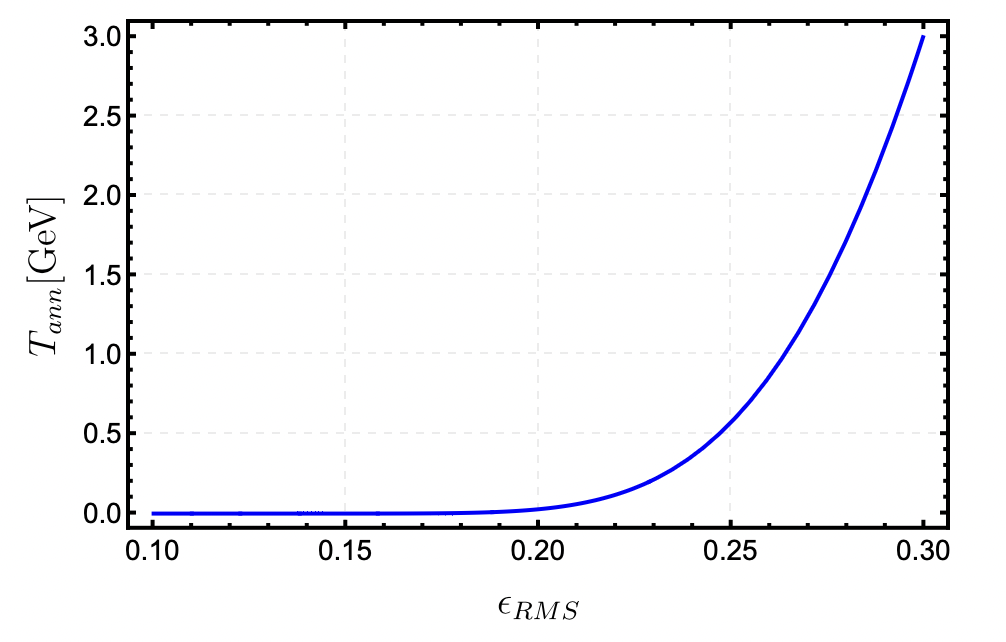}
    \caption{Impact of improving $D/H$ measurements on the bound in \eqref{cp even bound} with $T_{EW}\approx 130\text{ GeV}.$}
    \label{fig:Tann vs alpha}
\end{figure}

When the CP violation is even in the domain-wall field, the production of baryons during the scaling regime matters. Domain walls are expected to traverse the entire space and, since sphalerons are active within the domain walls, should lead to baryon production that covers the entire space. This, of course, occurs up to small inhomogeneities. If we say that a wall passes through a point $n$ times on average, then using the Poisson distribution we can estimate the amplitude of inhomogeneities to be $\sqrt{n}$. A density profile which reproduces the fractional error of $1/\sqrt{n}$ is given by
\begin{equation}
    \rho(\vec{x},0)=1+\frac{1}{\sqrt{n}}\cos\left(\frac{\pi x}{D}\right)\cos\left(\frac{\pi y}{D}\right),
\end{equation}
where $D$ is again the comoving Hubble radius and we again adopt a cosine profile for simplicity. We assume that the production due to the annihilation of domain walls is subdominant in this case. Evolving the initial baryon asymmetry profile with the diffusion equation and calculating the RMS fluctuations, we get the following constraint 
\begin{equation}
\label{eq:n-bound}
    n \gtrsim 2.78\ e^{-2d^2\pi^2/D^2}.
\end{equation}
Thus, if the walls on average traverse the volume three or more times, 
the bound will be satisfied.
What is the correct value of the  comoving scale $D$? For the case of the scaling solution, it must be somewhere between the DW separation in the beginning and the end of the baryogenesis process
\bea
\epsilon\ \eta|_{T=T_{i}}\leq D\leq \epsilon\ \eta|_{T=T_{f}}
\eea
where $T_{i}$ and $T_{f}$ are the temperatures when the baryogenesis starts and ends, respectively.
 Using the numbers from the previous section, we can rewrite the Eq.\eqref{eq:n-bound}  as 
\begin{equation}
    n \gtrsim 2.78 \ \exp\left[-\left(\frac{T_{*}}{0.82\text{ TeV}}\right)^2\right],\quad T_*\in [T_{i},T_{f}].
\end{equation}
Note that, by definition, $n\simeq V_{swept}/V_c$, 
where $V_{swept}$ is the comoving volume swept out by domain walls within a given comoving volume $V_c$. In \cite{Azzola:2024pzq} it has been estimated that 
\begin{equation}
\label{eq:n-def}
    n=\frac{V_{swept}}{V_c}\approx \frac{v}{\epsilon}\ln\left(\frac{T_i}{T_f}\right)+\frac{v}{\epsilon}\ln\left(\frac{T_{dw}-T_f}{T_{dw}-T_i}\right),
\end{equation}
where $T_{dw}$ is the temperature of domain-wall formation. Using the results from numerical simulations, namely that $\epsilon\approx 0.5$ and $\gamma v=0.4$, we obtain the following bound
\begin{equation}\label{bounddw}
    \ln\left[\frac{\left(\frac{T_{dw}}{T_f}\right)-1}{\left(\frac{T_{dw}}{T_i}\right)-1}\right]\gtrsim 3.76 \exp\left[-\left(\frac{T_{*}}{0.82\text{ TeV}}\right)^2\right].
\end{equation}
Using $T_i\approx T_{EW}\approx0.13$ TeV and $T_f\approx T_{ann}$ we can put bounds in the $(T_{ann},T_{dw})$ space. When the temperature at which the domain walls form is larger compared to the others, from \eqref{bounddw} we get
\begin{equation}
    \ln \left(\frac{T_{EW}}{T_{ann}}\right)\gtrsim 3.76 \exp\left[-\left(\frac{T_{*}}{0.82\text{ TeV}}\right)^2\right],
\end{equation}
which can be numerically solved to obtain (the result turns out to be approximately $T_*$ independent)
\begin{equation}\label{cp even bound}
    T_{ann}\lesssim 3 \text{ GeV}.
\end{equation}
Intuitively, this is sensible since the relative scale of inhomogeneities is roughly $1/\sqrt{n}$. Lower annihilation temperature makes the number of passings of domain walls greater, which tends to smooth out inhomogeneities. As before, this bound should be intersected with the requirement that the domain walls should annihilate before they start dominating the energy budget of the universe. This bound  was obtained using the constraint $\epsilon_{RMS}<0.3$. For the optimistic constraint $\epsilon_{RMS}<0.1$, we see that $T_{ann}$ is constrained to lie below BBN temperatures, which is incompatible with the predictions for light element abundances. Full dependence of $T_{ann}$ on $\epsilon_{RMS}$ is shown in Figure \ref{fig:Tann vs alpha}.\newline
\newline
The previous calculation explicitly assumed that the production of the baryon asymmetry is temperature-independent, which is not true in general. If the baryon asymmetry production depends on temperature, the magnitude of the resulting fluctuation may be estimated as follows:
\bea
\label{eq:neff}
&&\frac{1}{\sqrt n}\Rightarrow\frac{1}{\sqrt {n_{\rm eff}}}= \sqrt{\frac{\epsilon}{v}} \frac{\l[\int \frac{dT}{T} \eta^2(T)\r]^{1/2}}{\int \frac{dT}{T} \eta(T)},
\eea
where one can see that $n_{\rm eff}\to n$ (see Eq. \eqref{eq:n-def}) for constant $\eta(T)$. Then the bound from deuterium abundance becomes:
\bea
&&    n_{\rm eff} \gtrsim 2.78 \ \exp\left[-\left(\frac{T_{*}}{0.82\text{ TeV}}\right)^2\right],
\eea
where $T_*$ should be taken in the range with significant baryon asymmetry production. 
In Appendix \ref{Section: Toy model DWs}, we adopt a toy model of electroweak baryogenesis via domain walls, similar to the one in \cite{Azzola:2024pzq}, and 
study its temperature dependence. We find that the 
integral in Eq.\eqref{eq:neff} is dominated by a narrow temperature interval, yielding a small value, $n_{\rm eff} \approx 0.58$. Therefore,  if our findings remain valid for generic DW models with EW cores,
these models are excluded by the bounds from $D/H$ measurements.

\section{Summary and Conclusions}\label{Summary and Conclusions}
We conclude by summarizing the main results of the paper. We have reviewed the recent proposal to use $D/H$ constraints to put bounds on the baryon inhomogeneities in the early 
universe. In particular,  we have analysed the bounds on four classes of  baryogenesis  models with inhomogeneous baryon production and below we present a brief summary of the results for each of the models considered:
\bit
\item {\bf EW baryogenesis}: 
In most regions of the parameter space, the model remains consistent with the 
observed $D/H$ measurements. Intuitively, this is because, although 
the first-order phase transition is 
spatially inhomogeneous, baryon 
production stays approximately uniform 
throughout the bubble expansion process. 
The only changes to BAU generation arise 
from its dependence on temperature 
and bubble wall velocity; however, 
these quantities vary by at most $\mathcal{O}(1)$ factors over the course of the phase 
transition. In addition, constraints 
from primordial black hole production 
exclude scenarios with very large 
bubbles, of order the Hubble scale, 
further restricting the allowed 
variation in temperature and bubble 
velocity that could otherwise source BAU 
inhomogeneities.
\item {\bf Baryogenesis from the collisions of ultra-relativistic bubbles:} In this case, the BAU is produced only at the surfaces of bubble collisions, placing strong constraints on phase transitions with a characteristic scale below $100 \text{ GeV}$.
\item {\bf Baryogenesis from ultra-relativistic bubble-plasma collisions:} In this case, baryogenesis occurs from the heavy-particle production, which starts only after the bubbles have reached sufficient velocities. Thus, there will be empty regions corresponding to the initial expansion of the bubble. $D/H$ puts interesting constraints on the models with the phase transition scale below $\sim 100$ GeV. In this context, the bound depends on the mass of the heavy particle.

\item {\bf Baryogenesis from domain 
walls:} The constraints on this scenario are particularly strong because, during the scaling regime, the typical domain wall separation is of the order of the Hubble scale. If BAU production occurs primarily during domain wall annihilation, models with an annihilation scale below $\sim \mathrm{TeV}$ are ruled out. In models where BAU is generated mainly during the scaling regime, the bounds depend on the duration of that regime, becoming stronger as it becomes shorter. In particular, we find that the models where electroweak symmetry is restored in the core of the wall are excluded by $D/H$ measurements.
\eit
In summary, constraints on electroweak baryogenesis remain relatively weak, whereas more exotic scenarios are subject to tighter bounds. However, we stress that the set of models considered here is not exhaustive, and that additional scenarios may also be constrained with $D/H$ bounds. 
We leave a detailed investigation of these possibilities for future work.

\acknowledgments

We would like to thank  Giulio Barni
for helpful discussions and cross checks of the electroweak baryogenesis computations. 
This work was supported in part by the European Union - NextGenerationEU through the PRIN Project ``Charting unexplored avenues in Dark Matter'' (20224JR28W), and INFN initiative  APINE.


\appendix
\section{Diffusion lengths of protons and neutrons}\label{Appendix: Diffusion lengths for protons and neutrons}
In this Appendix, we review some of the results from \cite{Bagherian:2025puf} in calculating the diffusion lengths of protons and neutrons. The FLRW metric is given by
\begin{equation}
    ds^2=dt^2-a^2d\vec{x}^2,
\end{equation}
where $a$ is the scale factor and $\vec{x}$ are spatial comoving coordinates. The Hubble parameter in the radiation-dominated era is 
\begin{equation}
    H=\frac{\pi}{3}\frac{T^2}{M_{Pl}}\left(\frac{g_*(T)}{10}\right)^{1/2},
\end{equation}
where $M_{Pl}=2.4\cdot 10^{18}$ GeV is the reduced Planck mass. \newline
\newline
We are interested in determining the evolution of inhomogeneities in the baryon number density, i.e. to solve diffusion equations. Before neutrino decoupling, neutrons and protons are in equilibrium and we can treat their diffusion together. The diffusion will be dominated by the lowest temperatures because:
\begin{enumerate}
    \item the Hubble scale drops, allowing more time for diffusion ($t\sim 1/H$)
    \item interaction rates decrease, which leads to larger diffusion coefficients.
\end{enumerate}
At $T\sim 1 $ MeV, baryon diffusion is dominated by the scattering with electrons. The mean free path of neutrons is much smaller than that of protons, since they interact with electrons only via their magnetic dipole moment. Therefore, protons diffuse significantly only during the time they spent as neutrons, i.e. before neutrino decoupling. After neutrino decoupling, we must keep track of protons and neutrons separately. Since protons interact very strongly with electrons, their diffusion after this time is negligible.\newline
\newline
The diffusion equations for protons and neutrons are given by
\begin{equation}
    \begin{split}
        \partial_tn_n&=D_n\nabla^2n_n-\Gamma_{n\rightarrow p}n_n+\Gamma_{p\rightarrow n}n_p,\\
        \partial_tn_p&=D_p\nabla^2n_p-\Gamma_{p\rightarrow n}n_p+\Gamma_{n\rightarrow p}n_n,
    \end{split}
\end{equation}
where $\nabla^2=1/a^2 \partial_i \partial_i$ and $D_{n,p}$ are the diffusion coefficients. The rates consist of several processes
\begin{equation}
\begin{split}
    \Gamma_{n\rightarrow p}=\gamma_{ne^+\rightarrow p\bar{\nu}}+\gamma_{n\nu\rightarrow pe}+\gamma_{n\rightarrow pe\bar{\nu}}\\
    \Gamma_{p\rightarrow n}=\gamma_{p\bar{\nu}\rightarrow ne^+}+\gamma_{pe\rightarrow n\nu}+\gamma_{ pe\bar{\nu}\rightarrow n}\\
\end{split}
\end{equation}
where $\gamma_{x\rightarrow y}$ describes the thermally-averaged rate of the process $x\rightarrow y$. Detailed balance condition shows that \cite{Mukhanov:2005sc} (in the case $T\approx T_\nu$)
\begin{equation}
\begin{split}
    \gamma_{n\nu\rightarrow p e}&=\gamma_{p e\rightarrow n \nu}\exp(Q/T)\\
    \gamma_{ne^+\rightarrow p \bar{\nu}}&=\gamma_{p \bar{\nu}\rightarrow n e^+}\exp(Q/T),
\end{split}
\end{equation}
and $\gamma_{n\rightarrow p e \bar{\nu}}$ represents the neutron decay with lifetime $\tau_n=878$ seconds and the three-body rate $\gamma_{p e \bar{\nu}\rightarrow n}$ is negligible. The independent rates $ \gamma_{n\nu\rightarrow p e}$ and $\gamma_{ne^+\rightarrow p \bar{\nu}}$ can be found in \cite{Mukhanov:2005sc}. Using the fact that
\begin{equation}
    S\propto g_{*S}(T)T^3a(T)=const.
\end{equation}
we obtain the relation for the scale factor ($T_0=1$ MeV)
\begin{equation}
    a(T)=\frac{T_0}{T}\left(\frac{g_{*S}(T_0)}{g_{*S}(T)}\right)^{1/3}.
\end{equation}
It is easy to show that
\begin{equation}
    \frac{\partial}{\partial t}=-HT\left(1+\frac{1}{3}\frac{T}{g_{*S}(T)}\frac{\partial g_{*S}(T)}{\partial T}\right)^{-1}\frac{\partial}{\partial T},
\end{equation}
which gives the following expressions for the diffusion equations in momentum space
\begin{equation}
    \begin{split}
    -\frac{HT}{1+\frac{1}{3}\frac{T}{g_{*S}(T)}\frac{\partial g_{*S}(T)}{\partial T}}\frac{\partial \tilde{n}^{n}_k}{\partial T}&=\left(-D_nk^2\left(\frac{T}{T_0}\right)^2 \left(\frac{g_{*S}(T)}{g_{*S}(T_0)}\right)^{2/3}-\Gamma_{n\rightarrow p}\right)\tilde{n}^{n}_k+\Gamma_{p\rightarrow n}\tilde{n}^{p}_k,\\
    -\frac{HT}{1+\frac{1}{3}\frac{T}{g_{*S}(T)}\frac{\partial g_{*S}(T)}{\partial T}}\frac{\partial \tilde{n}^{p}_k}{\partial T}&=\left(-D_nk^2\left(\frac{T}{T_0}\right)^2 \left(\frac{g_{*S}(T)}{g_{*S}(T_0)}\right)^{2/3}-\Gamma_{n\rightarrow p}\right)\tilde{n}^{p}_k+\Gamma_{p\rightarrow n}\tilde{n}^{n}_k.
    \end{split}
\end{equation}
Neutron diffusion coefficients contain two contributions which combine as follows
\begin{equation}
    \frac{1}{D_n}=\frac{1}{D_{ne}}+\frac{1}{D_{np}},
\end{equation}
where ($x=T/m_e$)
\begin{equation}
    D_{ne}=\frac{\pi}{16}\left(\frac{m_n}{m_e}\right)^2\frac{1}{m_e(\alpha\kappa)^2}\frac{e^{1/x}}{x(1+3x+3x^2)}\approx 0.29 \frac{e^{1/x}}{x(1+3x+3x^2)} \text{cm} ,
\end{equation}
represents the interaction of the electron with the magnetic dipole moment of the neutron. The contribution $D_{np}$ comes from neutron-proton scattering and is negligible compared to the former. For the proton diffusion length, there is only a single contribution from proton-electron scattering given in \cite{Applegate:1987hm}
\begin{equation}
    D_p=\frac{3\pi}{8\alpha \Lambda(T)}\left(\frac{\hbar}{m_e}\right)\frac{x e^{1/x}}{1+2x+2x^2}\approx \frac{8.54 \cdot 10^{-7}}{\Lambda (T)} \frac{x e^{1/x}}{1+2x+2x^2} \text{cm},
\end{equation}
where $\Lambda(T)=\log (2/\theta_0)$ is the Coulomb logarithm which cuts-off the divergence at small scattering angles\footnote{The scattering  is cut-off by Debye screening at small scattering angles \cite{Bagherian:2025puf}. The Debye wave vector is 
given by $k_d^2=\frac{4\pi \alpha^2n_e}{a^3T}$ and the thermal wave vector is $k_{th}^2\sim3 m_eT$ in non-relativistic regime and $k_{th}^2\sim (3T)^2$ in the relativistic regime. The minimum angle is given by $\theta_0^2=k_d^2/k_{th}^2$.}. Although generally $T$-dependent, we adapt $\Lambda(T)\approx 5$ as in \cite{Applegate:1987hm}.
\begin{figure}[t!]
    \centering
    \subfloat[Diffusion lengths: neutrons (blue) and protons (orange)]{%
        \includegraphics[width=0.48\linewidth]{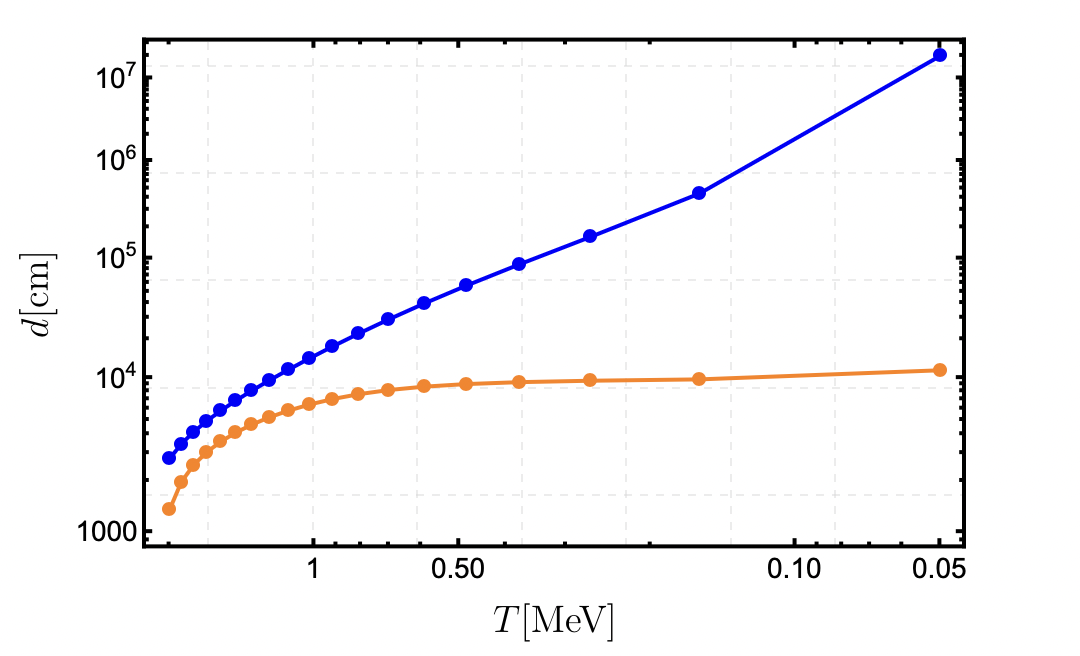}%
        \label{fig:diff length}%
    }
    \hfill
    \subfloat[Transfer function for protons (blue) and a Gaussian profile (orange) with the same diffusion length]{%
        \includegraphics[width=0.48\linewidth]{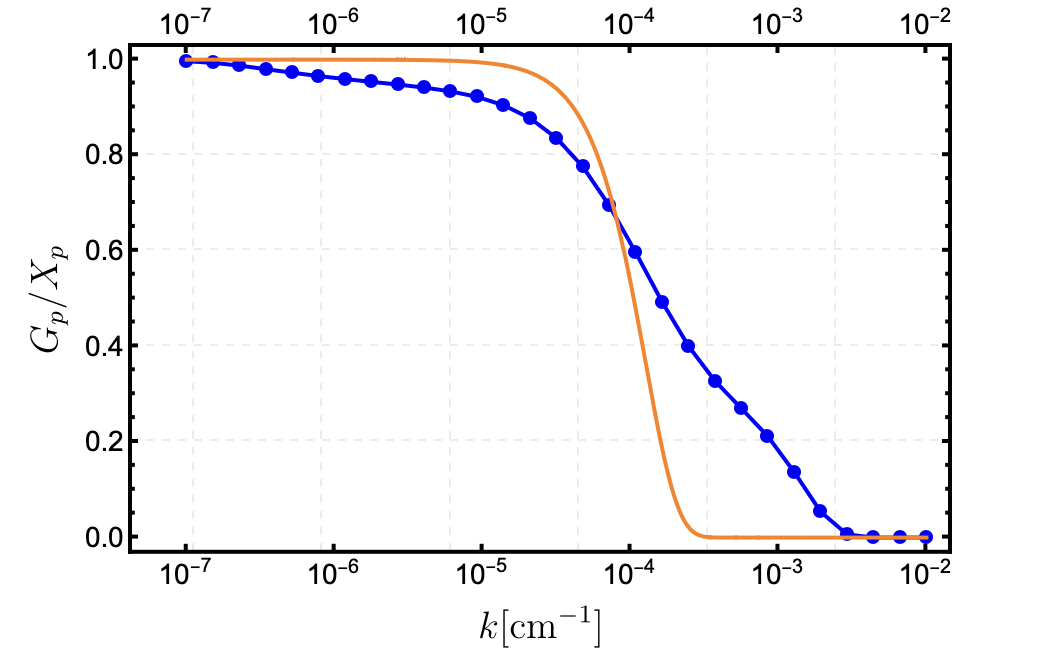}%
        \label{fig:transfer function}%
    }
    \caption{Diffusion lengths and the proton transfer function.}
    \label{fig:combined_diff_transfer}
\end{figure}
Since the equations are linear in number densities, the initial and final distributions are related by a transfer function. We ignore the initial division of baryons into $p$ and $n$, since for $T_i\gg$ MeV protons and neutrons are equilibrium. Implying that, for all $k$
\begin{equation}
\tilde{n}_k^p(T_i)=\tilde{n}_k^n(T_i)=\tilde{n}_k^B(T_i)
\end{equation}
where by $n_B$ we mean the number density of baryons. This means that
\begin{equation}
\tilde{n}_k^N(T)=\tilde{G}^N(k,T,T_i)\tilde{n}_k^B(T_i)
\end{equation}
where $N=n,p$. For $T_i$ such that $p$ and $n$ are in equilibrium, $\tilde{G}^N$ is insensitive to $T_i$ and we can drop it from the argument. We denote $X_N(T)=\tilde{G}^N(k=0,T)$. We define the diffusion length $d_i=1/k_0^i$ via the momentum scale
\begin{equation}
    \frac{\tilde{G}^{N}(k_0^i,T)}{X_N(T)}=\exp(-1/2).
\end{equation}
Numerical solution of the diffusion equations gives the results in Figure \ref{fig:diff length}. The proton transfer function deviates slightly from the Gaussian (Figure \ref{fig:transfer function}). This curve seems to be reasonably-fitted by
\begin{equation}
    G(k)\approx\exp\left[-\left(\frac{1}{7.14}\log_{10}(10^6d^2k^2)\right)^5\right].
\end{equation}
However, we observe that such a non-Gaussian profile changes the bounds up to order one numbers and for our purposes it is a good approximation to adopt a Gaussian profile with diffusion length $d_p$.
\section{Transport equations in EW baryogenesis}\label{Appendix: Transport equations}
In this Appendix, we describe the framework of the Boltzmann equations  used for electroweak baryogenesis (see for review \cite{Cline:2006ts}). First, we describe the formalism and then apply it to toy models of electroweak baryogenesis and electroweak baryogenesis via domain walls.
We will work in the WKB approximation, so that the wall profile varies slowly compared to the typical de Broglie wavelength of particles, namely
\begin{equation}
    l_w\gg \frac{1}{T}.
\end{equation}
The evolution of particles can then be separated into terms of different order in derivatives with respect to $z$. In particular, because of the $z$-dependence in the CP-violating phase $\Theta_t(z)$, there exists a CP-violating force proportional to the derivatives of $\Theta_t(z)$. Such a force will produce an excess of left-handed particles over antiparticles in front of the bubble wall. Since electroweak sphalerons are active in the symmetric phase, they efficiently convert left-handed particles to baryons. Once the baryons are produced, they are swept by the wall into the broken phase where the sphalerons are suppressed and the produced BAU cannot be washed out. We will closely follow \cite{Cline:2020jre,Barni:2025ifb} and while we leave all the technical details of the calculation to these papers, here we summarize the key points. \newline
\newline
We study the plasma of particles in the background of the bubble wall. Since the wall is moving, such a background drives the plasma out of equilibrium. In particular, we write the distribution of particle species $i$ as 
\begin{equation}
\begin{split}
    &f_i(z,\vec{p})=\frac{1}{\exp[\beta\gamma_w(E_0^i(z,\vec{p})+v_wp_z)-\beta\mu_i(z)]\pm1}+\delta f_i(z,\vec{p}), \\
    &E_{0i}=\sqrt{p_\bot^2+p_z^2+m^2_i(z)}, \quad \beta\equiv\frac{1}{T},
\end{split}
\end{equation}
where we parametrize the departures from chemical and kinetic equilibrium with $\mu_i$ and $\delta f_i$, respectively. That being said, $\delta f_i$ does not lead to a local number density fluctuation, namely
\begin{equation}
    \int d^3p \ \delta f_i(z,\vec{p})=0,
\end{equation}
and the fluctuation in number density is fully captured by $\mu_i(z)$. Transport equations are obtained starting from the Boltzmann equation
\begin{equation}\label{Boltzmann eqn}
    L[f]=(\dot{z}\partial_z+\dot{p}_z\partial_{p_z})f=C[f],
\end{equation}
where $L$ is the Liouville operator. The collision term $C[f]$ contains interactions amongst different particle species. One can split the fluctuations into CP-even and odd parts
\begin{equation}
    \mu=\mu_e+\mu_o, \quad \delta f=\delta f_e+\delta f_o,
\end{equation}
after which one can separate the Boltzmann equation into CP-even and CP-odd parts. As expected, the derivatives of $\Theta_t$ appear only in the CP-odd part. That being said, the excess of left-handed particles can only be generated from the CP-odd part in the Boltzmann equations, and we will focus on it in what follows. Additionally, we will drop the "odd" subscript in $\mu_o$ and $\delta f_o$.\newline
\newline
Once expanded up to second-order in fluctuations  and the gradient in $z$, the Boltzmann equation depends on three-momenta $\vec{p}$ of particle species, which makes it notoriously complicated to solve even numerically. To obtain a simpler description, the equations are projected onto a set of few moments, by averaging the Boltzmann equations over momentum with suitable weights, namely \cite{Cline:2020jre, Barni:2025ifb}
\begin{equation}
   \bigg \langle \left( \frac{p_z}{E_0}\right)^l L[f]\bigg\rangle = \bigg\langle \left( \frac{p_z}{E_0}\right)^l C[f]\bigg\rangle
\end{equation}
where $l=0,1,\dots$ and
\begin{equation}
    \langle X\rangle=\frac{1}{N_1} \int d^3p X_1(z,\vec{p}), \quad N_1\equiv\int d^3p  f^\prime_{0w,FD} (m=0)=-\gamma_w\frac{2\pi^3}{3}T^2 
\end{equation}
where
\begin{equation}
    f_{0w}(z,\vec{p})=\frac{1}{\exp[\beta\gamma_w(E_0^i(z,\vec{p})+v_wp_z))]\pm1}.
\end{equation}
We adopt the following convention for the derivative of $f_{0w}$\footnote{Note that later on we will also use $\prime=\frac{d}{dz}$. This notation has been used throughout the literature \cite{Cline:2020jre, Espinosa:2011eu, Barni:2025ifb,Bodeker:2004ws,Cline:2000nw} and we adopt it here for consistency.}
\begin{equation}
    f_{0w}^\prime(z,\vec{p})=\frac{\partial f(z,\vec{p})}{\partial (\gamma_w(E_0^i(z,\vec{p})+v_wp_z))}.
\end{equation}
The velocity moments are then given by
\begin{equation}\label{vel moments}
    u_l=\bigg\langle \left( \frac{p_z}{E_0}\right)^l \delta f (z,\vec{p})\bigg\rangle.
\end{equation}
We will restrict ourselves to the first two moments ($l=0$ and $l=1$) and obtain the two-equation fluid network
\begin{equation}\label{fluid network}
     A(z)w^\prime(z)+(m^2)^\prime B(z) w(z)=\mathcal{S}(z)+\delta \mathcal{C}(z),
\end{equation}
for each species and $w=(\mu(z) \ \ u(z))^T$ and in this equation $\prime=\frac{d}{dz}$. The vector $\mathcal{S}$ comes from the CP-odd sources in the Liouville operator and the vector $\delta \mathcal{C}$ comes from the collision terms. The exact expressions for $A$, $B$, $\mathcal{S}$ and $\delta \mathcal{C}$ are given below. This is a boundary value problem since we want the perturbations to vanish at infinity, i.e. $w(\pm\infty)=0$.\newline
\newline
The fluid network in \eqref{fluid network} describes the evolution of a single particle species, so we write this set of equations for each particle species under consideration. The different fluid networks are mixed through the collision terms. The particle species we will track are
\begin{equation}
    t_L: \ w_{t_L}=\begin{pmatrix}
    \mu_{t_L}\\
    u_{t_L}
    \end{pmatrix}, \quad b_L: \ w_{b_L}=\begin{pmatrix}
    \mu_{b_L}\\
    u_{b_L}
    \end{pmatrix}, \quad  t_R: \ w_{t_R}=\begin{pmatrix}
    \mu_{t_R}\\
    u_{t_R}
    \end{pmatrix}, \quad  h: \ w_{h}=\begin{pmatrix}
    \mu_{h}\\
    u_{h}
    \end{pmatrix}.
\end{equation}
Combining the different fluid networks gives us a set of eight first-order differential equations. Once the solution for the fluid network has been obtained, one can proceed to calculate the baryon asymmetry\footnote{Note that there is some discrepancy on the sign in front of the integral in the literature \cite{Barni:2025ifb,Cline:2020jre,vandeVis:2025efm}, but this is not important as one can always choose a different sign for the CP-violating phase to obtain $\mu_{B_L}\rightarrow-\mu_{B_L}$.}\cite{Cline:2020jre}
\begin{equation}\label{BAU equation}
    \eta_B=\frac{n_B}{s}=\frac{45n_fN_c}{4\pi^2v_w\gamma_wg_*T}\int_{-\infty}^\infty dz \ \mu_{B_L}(z)\Gamma_{ws}(z)\exp \left[-\frac{An_f}{2v_w\gamma_w}\Gamma_{ws} (z) \lvert z \rvert\right]
\end{equation}
where $n_f$ and $N_c$ are the number of generations and colors, respectively, and
\begin{equation}
\begin{split}
    g_\ast=106.75, 
    \quad \Gamma_{ws}(z)=\Gamma_{sph}f_{sph}(z), \quad \Gamma_{sph}=10^{-6}T, \quad f_{sph}(z)=\text{Min}\left[1,\frac{2.4T}{\Gamma_{sph}}e^{-40v(z)/T}\right]
\end{split}
\end{equation}
and 
\begin{equation}\label{baryon chem pot}
    \mu_{B_L}(z)=\frac{1}{2}\sum_{\text{quarks}}\mu_q(z)=\frac{1}{2}(1+4D_0^t)\mu_{t_L}(z)+\frac{1}{2}(1+4D_0^b)\mu_{b_L}(z)+2D_0^t\mu_{t_R}(z).
\end{equation}
It is convenient to solve the fluid network equations in terms of dimensionless variables: $\tilde{w}=w/T$ and $\tilde{z}=zT$, introducing  $x=m/T$ \footnote{Note that the velocity moments \eqref{vel moments} are defined as dimensionful quantities with mass dimension one.}.  In the following, we give exact expressions of the various terms in the Boltzmann equation.
\subsection{Moments of the fluid network}
As already mentioned, we adopt the conventions from \cite{Cline:2020jre,Barni:2025ifb} and the reader can consult these papers for a detailed derivation of the transport equations. We write again the transport equations given in \eqref{fluid network}
\begin{equation}
     A(z)w^\prime(z)+(m^2)^\prime B(z) w(z)=\mathcal{S}(z)+\delta \mathcal{C}(z),
\end{equation}
where 
\begin{equation}
    A=\begin{pmatrix}
        -D_1 & 1\\
        -D_2 &-v_w
    \end{pmatrix} \quad B=\begin{pmatrix}
        v_w\gamma_w Q_1 & 0\\
        v_w\gamma_w Q_2 & \bar{R}
    \end{pmatrix}
\end{equation}
and 
\begin{equation}
    D_l=\bigg \langle \left( \frac{p_z}{E_0}\right)^l f^\prime_{0w}\bigg\rangle, \quad Q_l=\bigg \langle \frac{p_z^{l-1}}{2E_0^{l}} f^{\prime\prime}_{0w}\bigg\rangle
\end{equation}
and 
\begin{equation}
    \bar{R}=\frac{\pi}{\gamma_w^2\hat{N_0}}\int_m^\infty dE\log\left\lvert\frac{p-v_w E}{p+v_w E}\right\rvert f_0(E), \quad p=\sqrt{E^2-m^2}
\end{equation}
where $f_0(E)$ is the isotropic distribution in the plasma frame and 
\begin{equation}
    N_0=\int d^3 p\ f_{0w}(m)=\gamma_w\hat{N_0}.
\end{equation}
\subsubsection{Collision terms}
Collision terms $\delta \mathcal{C}$ describe how interactions relax the system from small departures from equilibrium and we divide them into \cite{Cline:2020jre,Barni:2025ifb}:
\begin{itemize}
    \item inelastic terms (number changing): they drive some linear combination of $\mu_i$ towards zero
    \item elastic terms (number conserving): they relax the momentum space distribution $\delta f$
\end{itemize}
Once the collision terms have been written one can perform the projection onto the moments of the Boltzmann equation and obtain the general linearised collision operator\footnote{For details on the derivation of this expression see \cite{Barni:2025ifb}.}
\begin{equation}\label{generalised collision term}
    \delta\mathcal{C}_{l+1}^{a}=-K_{a}^{(l)}\sum_{i,j}\hat{\Gamma}^{\text{inel}}_{i\rightarrow j}s_{ij}\mu_j-K_a^{(0)}\delta_{l0}\Gamma_{ss}\mu_{ss}-\hat{\Gamma}_{tot}^a u_{l,a}.
\end{equation}
where
\begin{equation}
    K_a^l=\bigg \langle \left(\frac{p_z}{E_0}\right)^l f^a_{0w}\bigg\rangle
\end{equation}
and $s_{ij} \in \{-1,0,1\}$ depending on how the particle species enters the interaction (negative if it is in the initial state and positive if it is in the final state). It is important to explain various terms in Eq. \eqref{generalised collision term} 
\begin{itemize}
    \item $\hat{\Gamma}^{\text{inel}}_{i\rightarrow j}$ represents the rate of various interactions which we will include (Yukawa, charged-current etc.).
    \item The second term in \eqref{generalised collision term} is due to strong sphalerons, such a term is generally represented in the collision term as
    \begin{equation}
        \delta \mathcal{C}\supset -c_a^{ss}\Gamma_{ss}\sum_{\text{quarks}} (\mu_{q_L}-\mu_{q_R}).
    \end{equation}
    Since light quark Yukawa interactions are slow, light quarks only enter the fluid network through strong sphalerons. Looking closely at the fluid network of light quarks, one can notice that their chemical potentials satisfy $\mu_{q_L}=-\mu_{q_R}$ and that it can be fully written in terms of heavy species. Then one obtains
    \begin{equation}
        \sum_{\text{quarks}}(\mu_{q_L}-\mu_{q_R})=(1+9D_0^t)\mu_{q_L}+(1+9D_0^b)\mu_{b_L}-(1-9D_0^t)\mu_{t_R}\equiv\mu_{ss}.
    \end{equation}
    It is a reasonable approximation to restrict only to the lowest modes of strong sphaleron collision term \cite{Kainulainen:2024qpm}.
    \item The $\hat{\Gamma}_{tot}$ acts as a damping term for higher-velocity moments ($u_{0a}=0$ by definition) and defines a relaxation of momentum perturbations due to the collision with the plasma. It is useful to recast it in terms of moment functions
    \begin{equation}
        \hat{\Gamma}_{tot}^a=-\frac{D_{2,a}v_w}{D_{1,a} D^a_{diff}},
    \end{equation}
    where $D^a_{diff}$ is the physical diffusion constant of some species $a$.
\end{itemize}
The lowest moment collision terms which we will add to our network are
\begin{equation}
    \begin{split}
        &\delta\mathcal{C}_1^{t_L}=-K_t^{(0)}\left[\hat{\Gamma}^t_y(\mu_{t_L}-\mu_{t_R})+2\hat{\Gamma}_m^t(\mu_{t_L}-\mu_{t_R})+\hat{\Gamma}_W(\mu_{t_L}-\mu_{b_L})+\Gamma_{ss}\mu_{ss}\right],\\
        &\delta\mathcal{C}_1^{b_L}=-K_b^{(0)}\left[\hat{\Gamma}^t_y(\mu_{b_L}-\mu_{t_R})+\hat{\Gamma}_W(\mu_{b_L}-\mu_{t_L})+\Gamma_{ss}\mu_{ss}\right],\\
        &\delta\mathcal{C}_1^{t_R}=-K_t^{(0)}\left[-\hat{\Gamma}^t_y(\mu_{t_L}+\mu_{b_L}-2\mu_{t_R})+2\hat{\Gamma}_m^t(\mu_{t_R}-\mu_{t_L})-\Gamma_{ss}\mu_{ss}\right],\\
         &\delta\mathcal{C}_1^{h}=-K_h  ^{(0)}\left[\frac{3}{2}\hat{\Gamma}^t_y(\mu_{t_L}+\mu_{b_L}-2\mu_{t_R}+2\mu_h)+\hat{\Gamma}_m^h\mu_h\right],\\
        &\delta\mathcal{C}_2^{t_L}=-K_t^{(1)}\left[\hat{\Gamma}^t_y(\mu_{t_L}-\mu_{t_R})+2\hat{\Gamma}_m^t(\mu_{t_L}-\mu_{t_R})+\hat{\Gamma}_W(\mu_{t_L}-\mu_{b_L})\right]-\hat{\Gamma}_{tot}^tu_{1,t_L},\\
        &\delta\mathcal{C}_2^{b_L}=-K_b^{(1)}\left[\hat{\Gamma}^t_y(\mu_{b_L}-\mu_{t_R})+\hat{\Gamma}_W(\mu_{b_L}-\mu_{t_L})\right]-\hat{\Gamma}_{tot}^bu_{1,b_L},\\
        &\delta\mathcal{C}_2^{t_R}=-K_t^{(1)}\left[-\hat{\Gamma}^t_y(\mu_{t_L}+\mu_{b_L}-2\mu_{t_R})+2\hat{\Gamma}_m^t(\mu_{t_R}-\mu_{t_L})\right]-\hat{\Gamma}_{tot}^tu_{1,t_R}.\\
        &\delta\mathcal{C}_2^{h}=-K_h  ^{(1)}\left[\frac{3}{2}\hat{\Gamma}^t_y(\mu_{t_L}+\mu_{b_L}-2\mu_{t_R}+2\mu_h)+\hat{\Gamma}_m^h\mu_h\right]-\hat{\Gamma}_{tot}^h u_{1,h}
    \end{split}
\end{equation}
where $\hat{\Gamma}_y$ represents the rate due to Yukawa interactions, $\hat{\Gamma}_W$ represents the charged current interactions and $\hat{\Gamma}_m$ represents the helicity-flip processes $t_L\leftrightarrow t_R$ from the interaction with the wall. For the rates and diffusion constants we take \cite{Cline:2020jre, Fromme:2006wx}
\begin{equation}
\begin{split}
    &\hat{\Gamma}_y^t\approx 4.2 \times 10^{-3}T, \quad \hat{\Gamma}_W\approx\hat{\Gamma}^h_{tot}, \quad \hat{\Gamma}^t_m\approx\frac{m_t^2}{63T}, \quad \hat{\Gamma}_m^h\approx \frac{m_W^2}{50T},\\
    &\Gamma_{ss}\approx4.9\times 10^{-4}T, \quad D_{q}\approx\frac{6}{T}, \quad D_h\approx\frac{20}{T}.
\end{split}
\end{equation}
Note that the weak sphalerons are treated as slow compared to the other processes in the fluid network and are only turned on later, when we compute the baryon asymmetry.
\subsubsection{Sources}
In this section, we summarize how to compute weighted sources for our fluid network. We note again that we only restrict ourselves to the CP-odd part. This is characterized by two moments
\begin{equation}
    Q_l^{8o}=\bigg\langle \frac{s p_z^{l-1}}{2E_0^lE_{0z}}f_{0w}^\prime\bigg\rangle, \quad Q_l^{9o}=\bigg\langle\frac{p_z^{l-1}}{4E_0^{l+1}E_{0z}}\left(\frac{f_{0w}^\prime}{E_0}-\gamma_wf^{\prime\prime}_{0w}\right)\bigg\rangle,
\end{equation}
where $E_{0z}=\sqrt{p_z^2+m^2}$. With these definitions, the CP-odd source becomes
\begin{equation}
    \mathcal{S}_l^{odd}(z)=-hv_w\gamma_w\left[(m^2\Theta^\prime)^\prime Q_l^{8o}-m^2(m^2)^\prime\Theta^\prime Q_l^{9o}\right],
\end{equation}
where with $h=\pm 1$ we track the helicity-dependence of the source.
\subsubsection{Useful integrals}
The integrals which compute the various moments can be very difficult to implement even numerically, and hence we use the simplifications in \cite{Barni:2025ifb,Cline:2020jre} to reduce them to dimensionless integrals. Namely, we use dimensionless plasma frame variables
\begin{equation}
    w\equiv\frac{E}{T}, \quad x\equiv\frac{m}{T}, \quad \tilde{p}_w=\sqrt{w^2-x^2}, \quad y=\cos\theta=\frac{p_z}{\lvert \vec{p}\rvert}.
\end{equation}
Using the Lorentz-invariance of the integration measure, we can write all the integrals in terms of a dimensionless one
\begin{equation}
    \bigg\langle\frac{p_z^n}{E^m}V\mathcal{F}^{(k)}_{0w}\bigg\rangle= T^{n-m-k-1}K(\mathcal{F}_0^{(k)};V,n,m)
\end{equation}
where $\mathcal{F}_{0w}^{(k)}=f_{0w}^{(k)}$, $\mathcal{F}_0^{(k)}=f_0^{(k)}$ and $f_0(E)$ is the isotropic distribution in the plasma frame. The dimensionless integral is given by
\begin{equation}
    K(\mathcal{F}_0;V,n,m)=-\frac{3}{\pi^2\gamma_w}\int_x^\infty dw \int_{-1}^1dy \frac{\tilde{p}_w\tilde{p}_z^{n}}{\tilde{E}^{m-1}}V(w,y)\mathcal{F}_0(w),
\end{equation}
where 
\begin{equation}
    \tilde{p}_z=\gamma_w(y\tilde{p}_w-wv_w), \quad \tilde{E}=\gamma_w(w-v_wy\tilde{p}_w).
\end{equation}
For the moments $D_l, Q_l, K_l$ the function $V=1$, whereas for the CP-odd source terms $V$ is non-trivial. In particular, in the spin basis
\begin{equation}
    V=V_s=\frac{\lvert \tilde{p_z}\rvert}{\sqrt{\tilde{p}^2_z+x^2}},
\end{equation}
or in the helicity basis
\begin{equation}
    V_h=V_s^2\left(1-\frac{x^2}{w^2}\right)^{-1/2}.
\end{equation}
Different basis choices however lead to small differences in source terms \cite{Cline:2020jre,Cline:2017qpe}. Thus, the different moments in the fluid network can be written as 
\begin{equation}
    \begin{split}
        D_l&=K(f^\prime_{0w};1,l,l), \\
        T^{-1}K_l&=K(f_{0w};1,l,l), \\
        T^2Q_l&=\frac{1}{2}K(f^{\prime\prime}_{0w};1,l-1,l),\\
        T^2Q_l^{8o}&=\frac{1}{2}K(f^{\prime}_{0w};V,l-2,l),\\
        T^4Q_l^{9o}&=\frac{1}{4}\left[K(f^{\prime}_{0w};V,l-2,l+2)-\gamma_wK(f^{\prime\prime}_{0w};V,l-2,l+1)\right].
    \end{split}
\end{equation}
\begin{figure}[t!]
    \centering
    \subfloat[Different chemical potentials $\mu_{t_L}/T, \mu_{b_L}/T, \mu_{t_R}/T$ and $\mu_h/T$.]{
        \includegraphics[width=0.45\textwidth]{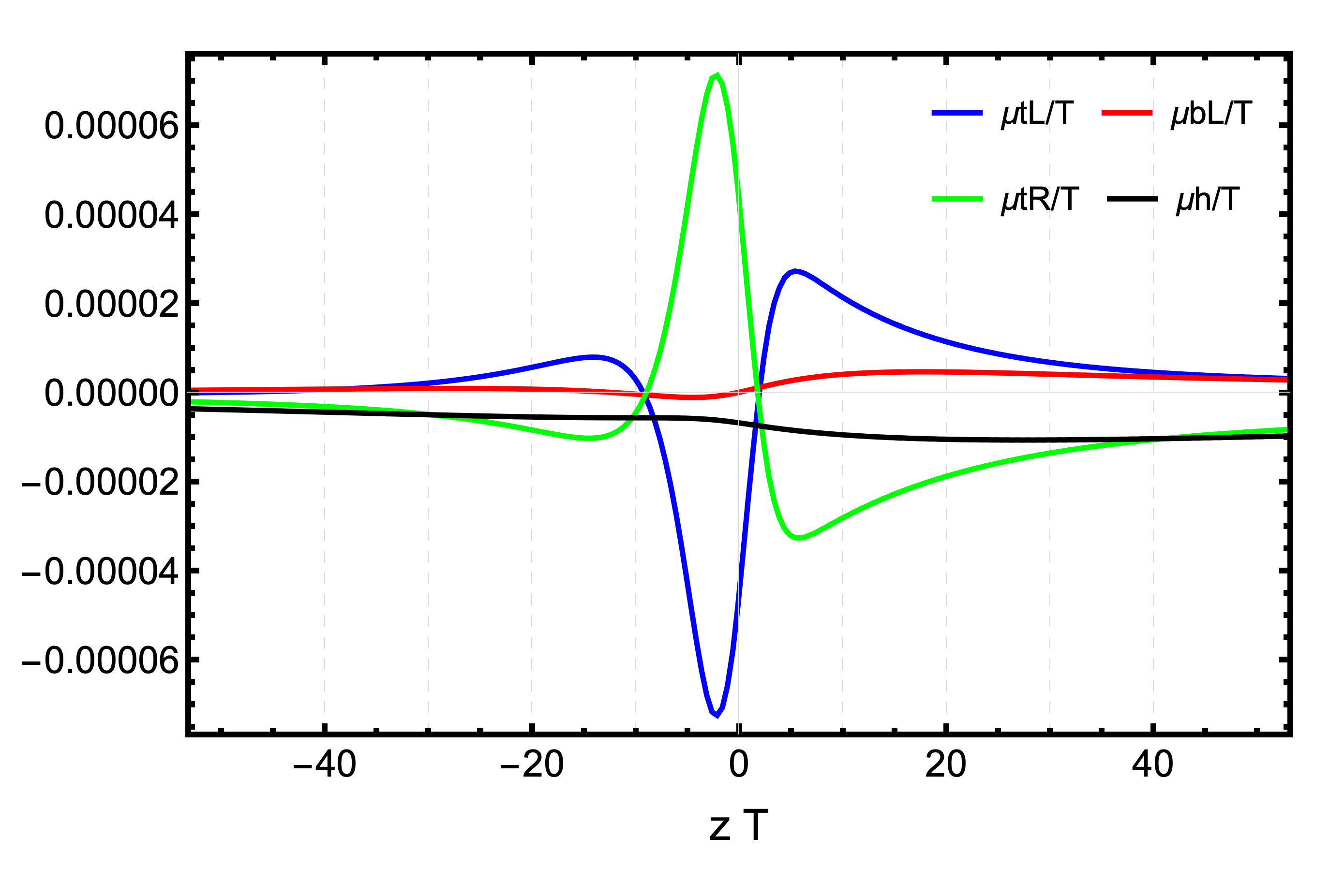}
        \label{fig:mutlbltr}
        }
    \hfill
    \subfloat[Baryon-left chemical potential $\mu_{B_L}/T$.]{
        \includegraphics[width=0.45\textwidth]{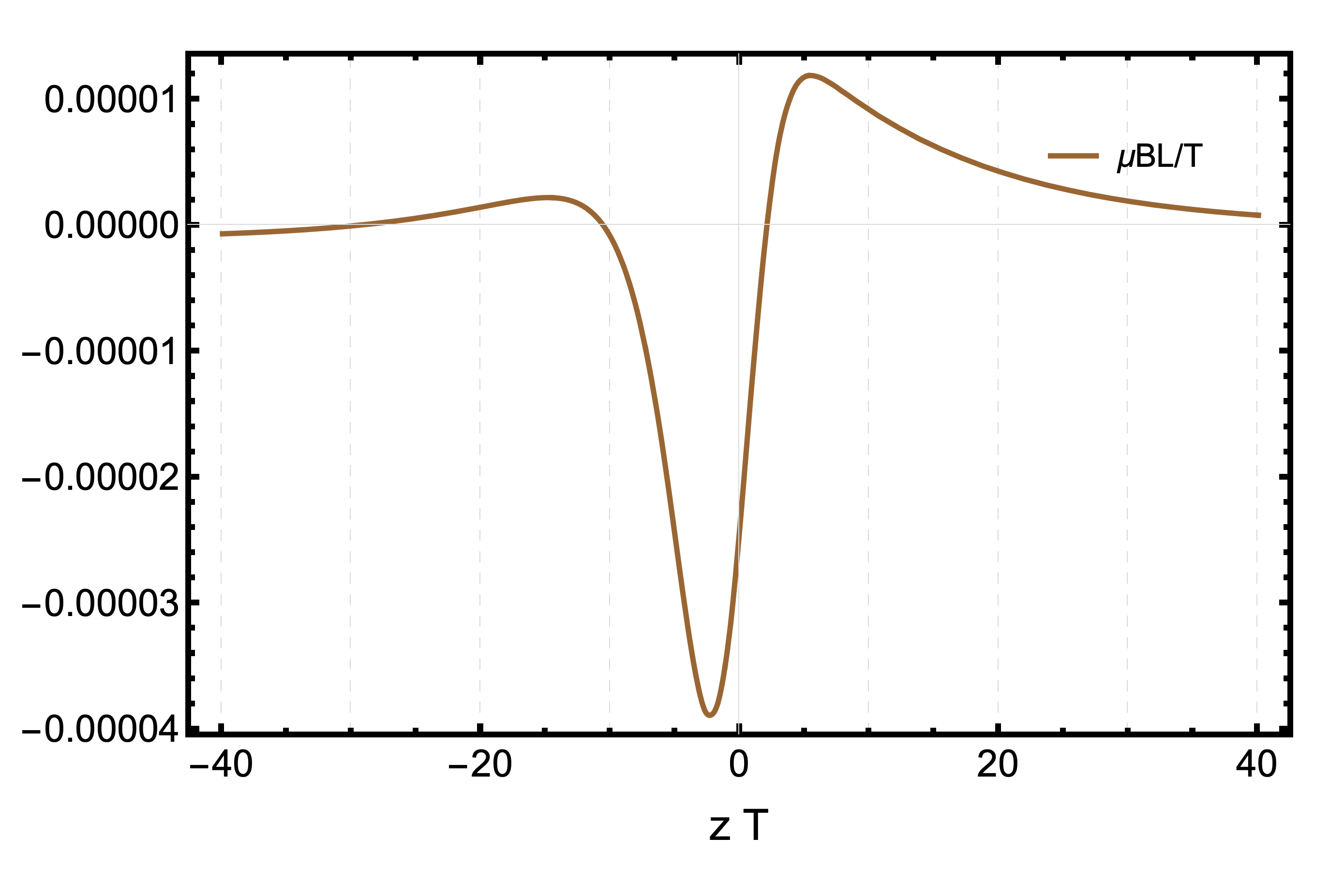}
        \label{fig:muBl}
    }
    \caption{Solutions obtained from the fluid network for $v_c=T$, $l_w T=5$, $v_w=0.1$, $y_t=0.7$ and $s_n/\Lambda=0.2$.}
    \label{fig:Transport eqns chemical potentail}
\end{figure}
\subsection{EW baryogenesis: a toy model}\label{Section: Toy model Sm with a singlet}
\begin{figure}[t!]
    \centering
    \subfloat[Temperature dependence of $\eta_B/\eta_{B}^{exp}$ for different wall widths.]{
        \includegraphics[width=0.4\linewidth]{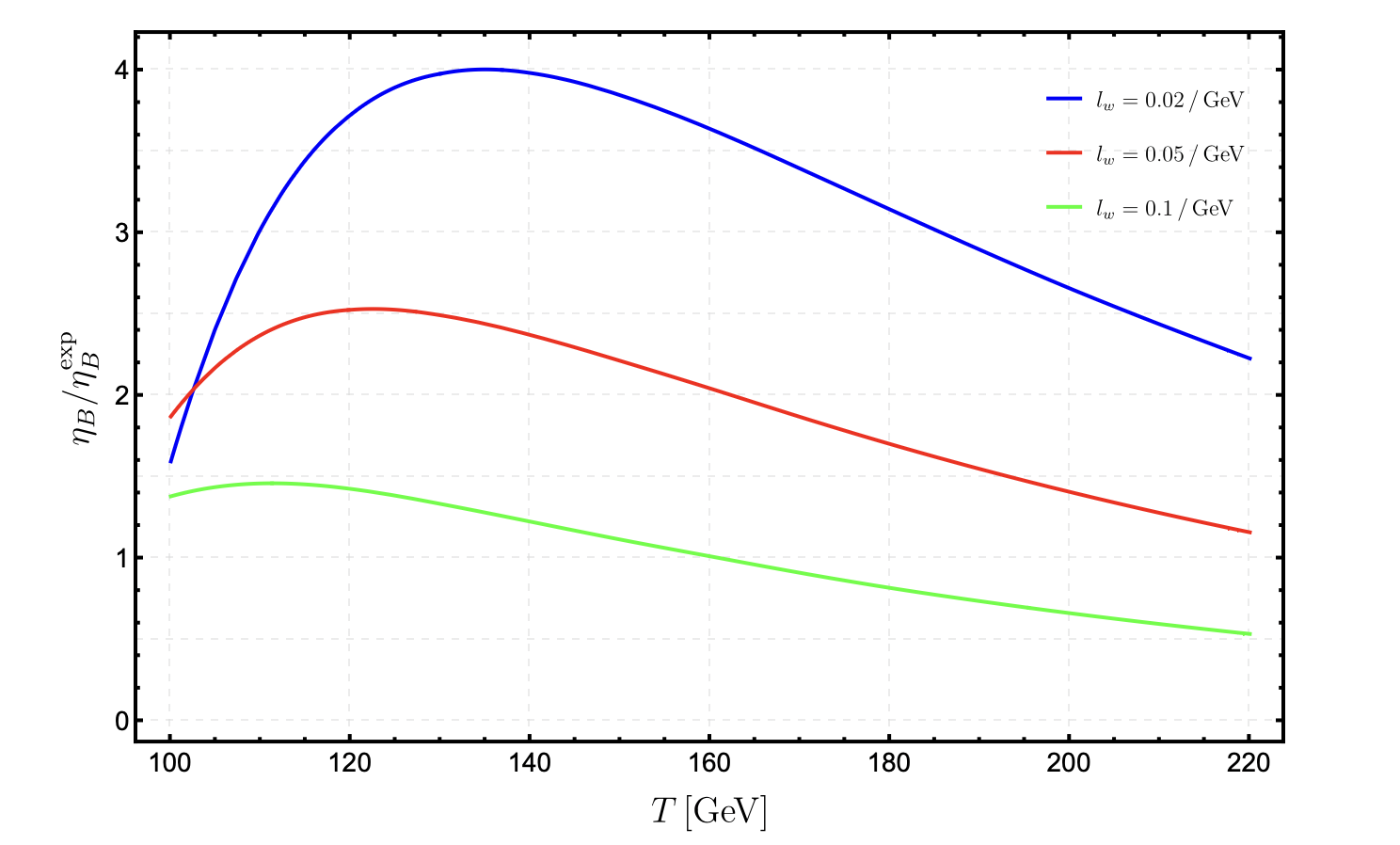}
        \label{fig:tdependence}
    }
    \hfill 
    \subfloat[The derivative $|d \log\eta(T)/d\log T|$ for various wall widths.]{
        \includegraphics[width=0.4\linewidth]{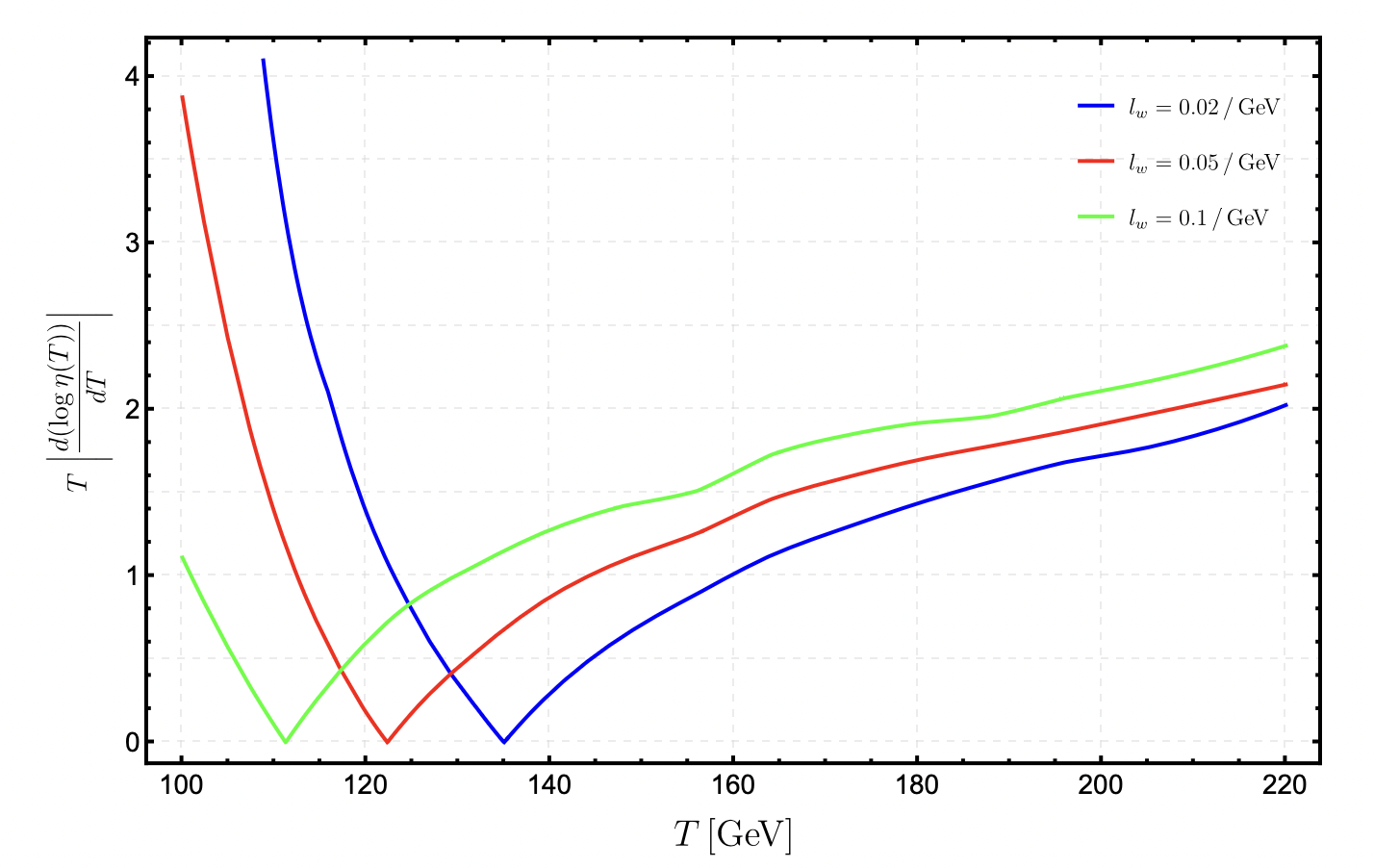}
        \label{fig:dlogchidT}
    }
    
    \caption{Temperature dependence of EW baryogenesis for $v_{EW}=246$ GeV, $v_w=0.1$, $s_n/\Lambda=0.2$ and $y_t=0.7$.}
    \label{fig:combined_plots}
\end{figure}

\begin{figure}[t!]
    \centering
    \subfloat[$l_w=0.02\text{ GeV}^{-1}$]{
        \includegraphics[width=0.4\linewidth]{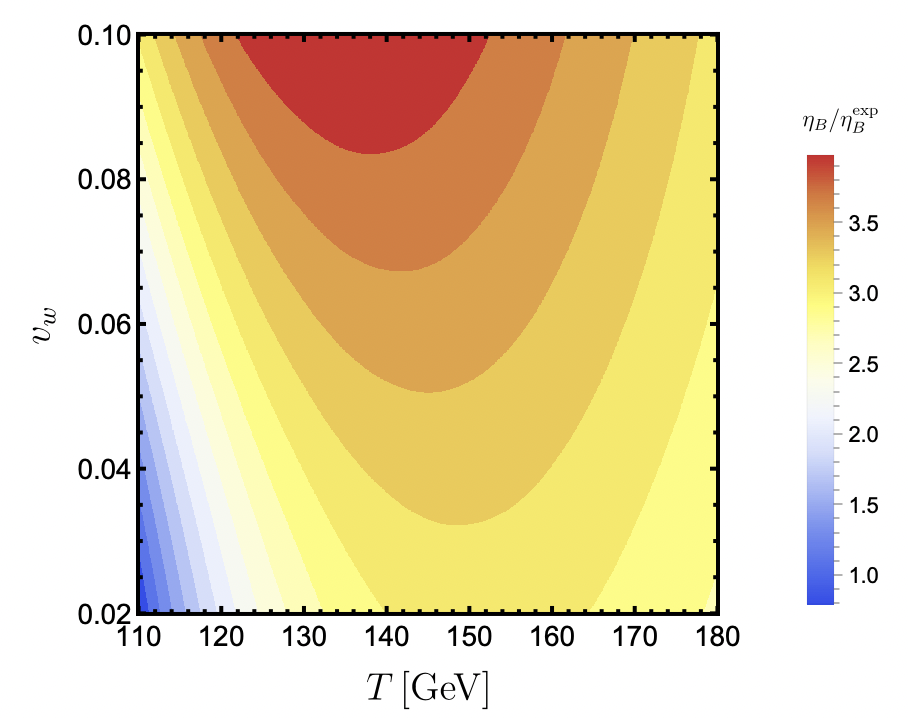}
        \label{fig:EWBGcontour1}
    }
    \hfill
    \subfloat[$l_w=2/T$]{
        \includegraphics[width=0.38\linewidth]{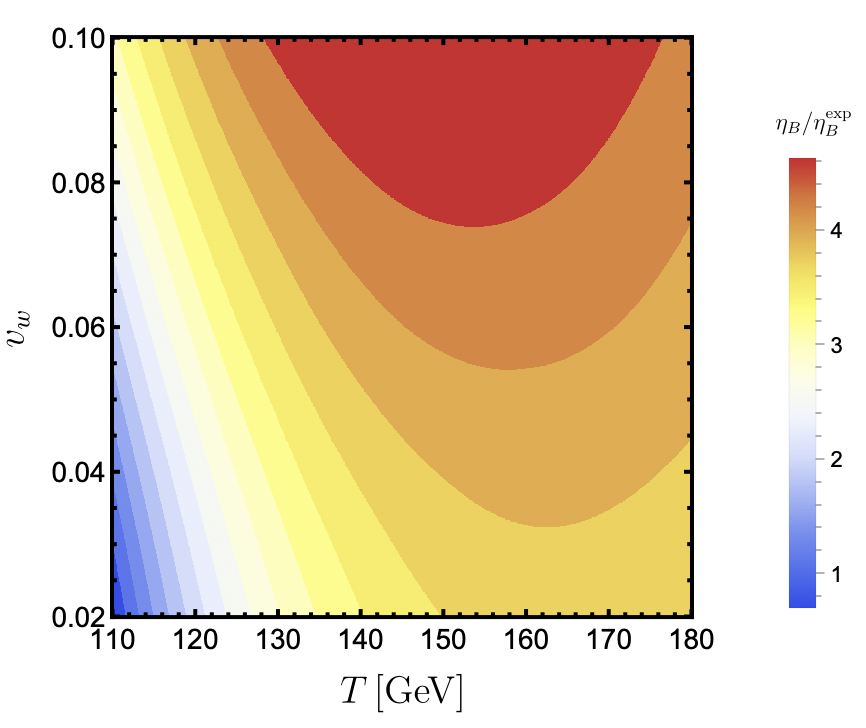}
        \label{fig:EWBGcontour2}
        }
    \caption{Contour plot of $\eta_B/\eta_B^{exp}$ in the $(v_w,T)$ plane for different wall width choices. The rest of the parameters are set as before: $v_{EW}=246$ GeV, $s_n/\Lambda=0.2$ and $y_t=0.7$.}
    \label{fig:EWBGcontours}
\end{figure}

We assume that the electroweak phase transition is of first-order due to the effects of physics beyond the Standard Model. We focus on the motion of the bubble wall when its radius is sufficiently large so that it can be locally treated as planar, and the problem becomes one-dimensional. We will work in the rest frame of the wall and assume that $z$ axis is perpendicular to the wall. The bubble-wall profile in the $z$-direction is given by
\begin{equation}
    v(z)=\frac{v_{c}}{2} \left[1-\tanh\left(\frac{z}{l_w}\right)\right],
\end{equation}
where $v_c$ is the Higgs VEV and $l_w$ is the wall-width. CP violation arises from the presence of a singlet scalar field that couples to the top quark through a dimension-five operator
\cite{Espinosa:2011eu,Barni:2025ifb}
\begin{equation}
    \mathcal{L}\supset y_t v(z) \bar{t}_L \left(1+i\frac{s(z)}{\Lambda}\right)t_R+\text{h.c.}.
\end{equation}
Similar interactions will be present for other fermions; however, CP violation in the top quark sector is expected to dominate, since it has the largest Yukawa coupling.
We assume that the singlet has the wall-profile opposite to the Higgs
\begin{equation}
s(z)=\frac{s_n}{2}\left[1+\tanh\left(\frac{z}{l_w}\right)\right],
\end{equation}
so that in the symmetric phase $\langle s\rangle =s_n$ and in the broken $\langle s \rangle=0$\footnote{Note that we made the usual assumption that the wall widths in $v(z)$ and $s(z)$ are the same, which generally does not need to be the case \cite{Barni:2025ifb}.}. The effective complex top mass is given by
\begin{equation}
    m_t(z)=\lvert m_t(z)\rvert e^{i\Theta_t(z)},
\end{equation}
where
\begin{equation}
    m_t(z)=y_tv(z) \sqrt{1+\frac{s(z)^2}{\Lambda^2}}, \quad \Theta_t(z)=\arctan\left(\frac{s(z)}{\Lambda}\right).
\end{equation}
Typical curves one gets for the chemical potentials are shown in Figure \ref{fig:Transport eqns chemical potentail}. We show the temperature dependence in Figure \ref{fig:tdependence}, by fixing $v_{EW}=246$ GeV and $v_w=0.1$, for different wall widths\footnote{Note that we did not include the Higgs while studying this temperature dependence. Inclusion of the Higgs leads to numerical instability and requires much larger numerical precision, whereas at the same time its effect on the final baryon asymmetry is small.}.
We remind the reader that this approximation is valid only in the limit $l_w\gg 1/T$, which is why we only investigated the temperature dependence down to $\sim 100$ GeV for the wall widths above. Note that we must have that $v_{EW}/T\gtrsim 1$, so that sphalerons are suppressed in the broken phase. As discussed in Section \ref{Section: Electroweak baryogenesis}, we are interested in the derivative of the logarithm of the function in Figure \ref{fig:tdependence}. We find that
\begin{equation}
  T\left\lvert\frac{d}{dT}\left[\ln\left(\frac{\eta_B(T)}{\eta_{B}^{exp}}\right)\right]\right\rvert\sim \mathcal{O}(1),
\end{equation}
and we sketch this function for various wall widths in Figure \ref{fig:dlogchidT}.\newline
\newline
We also give contour plots of $\eta_B/\eta_B^{exp}$ in the ($v_w,T$) plane for different choices of wall width in Figure \ref{fig:EWBGcontours}. Since the wall velocity is generally expected to be temperature dependent, we can obtain the maximal values of the logarithimic derivative for various temperature dependences of the velocity. Namely, we get
\begin{equation}
\begin{split}
    v_w\sim \frac{1}{T}&: \quad \text{Max}\left \lvert\frac{d\log \eta}{d\log T}\right\rvert\approx 2.71,\\   v_w\sim\frac{1}{T^2}&: \quad \text{Max}\left \lvert\frac{d\log \eta}{d\log T}\right\rvert\approx 1.82,\\ v_w\sim\frac{1}{T^3}&: \quad \text{Max}\left \lvert\frac{d\log \eta}{d\log T}\right\rvert\approx 1.18,
\end{split}
\end{equation}
where we normalized the velocity to $v_w(T=110\text{ GeV})=0.1$. We see that even by assuming some temperature dependence of the velocity, we cannot achieve very strong $\eta$ dependence on temperature.
\newline

\subsection{EW baryogenesis via domain walls: a toy model}\label{Section: Toy model DWs}
A similar analysis applies to BAU generation in the model of Ref. 
\cite{Azzola:2024pzq}. In this framework, the BAU is generated after electroweak 
symmetry breaking through the evolution of a domain wall network. The model 
predicts the existence of domain walls whose cores preserve electroweak 
symmetry, thereby allowing baryon number violation to proceed via sphaleron processes localized within the wall cores. In addition to the SM, the model contains
a real singlet scalar field $s$ with an 
approximate $Z_2$ symmetry, which is 
broken spontaneously leading to the 
appearance of domain walls. In this setup, we would like to estimate  the  values of the quantity $n_{\rm eff}$ introduced in the Section \ref{Section: DW scaling regime}.\newline
\newline
We will assume that  CP violation  comes from the following dimension six operator with top quarks:
\begin{equation}
    \mathcal{L}\supset y_t v(z) \bar{t}_L \left[1+i\left(\frac{s(z)}{\Lambda}\right)^2\right]t_R+\text{h.c.}
\end{equation}
The effective top mass in this case is given by
\begin{equation}
    m_t(z)=\lvert m_t(z)\rvert e^{i\Theta_t(z)},
\end{equation}
where
\begin{equation}
    m_t(z)=y_tv(z) \sqrt{1+\left(\frac{s(z)}{\Lambda}\right)^4}, \quad \Theta_t(z)=\arctan\left[\left(\frac{s(z)}{\Lambda}\right)^2\right].
\end{equation}
For the Higgs profile $v(z)$ we adopt the model of \cite{Azzola:2024pzq}, where the electroweak symmetry is restored in small regions referred to as the electroweak cores. We model the elecroweak cores with the following profile
\begin{equation}
    v(z)=v_{EW}\left[1+\frac{1}{2}\tanh\left(\frac{z-z_c}{l_w}\right)-\frac{1}{2}\tanh\left(\frac{z+z_c}{l_w}\right)\right],
\end{equation}
where $z_c$ is determined such that at $v(0)=h_0$, hence
\begin{equation}
    z_c=l_w\text{arctanh}\left(1-\frac{h_0}{v_{EW}}\right).
\end{equation}
\begin{figure}[t!]
    \centering
    \subfloat[Profiles for $v(z)$ and $s(z)$ where $l_wT=40$ and $h_0=50$ GeV.]{
        \includegraphics[width=0.45\linewidth]{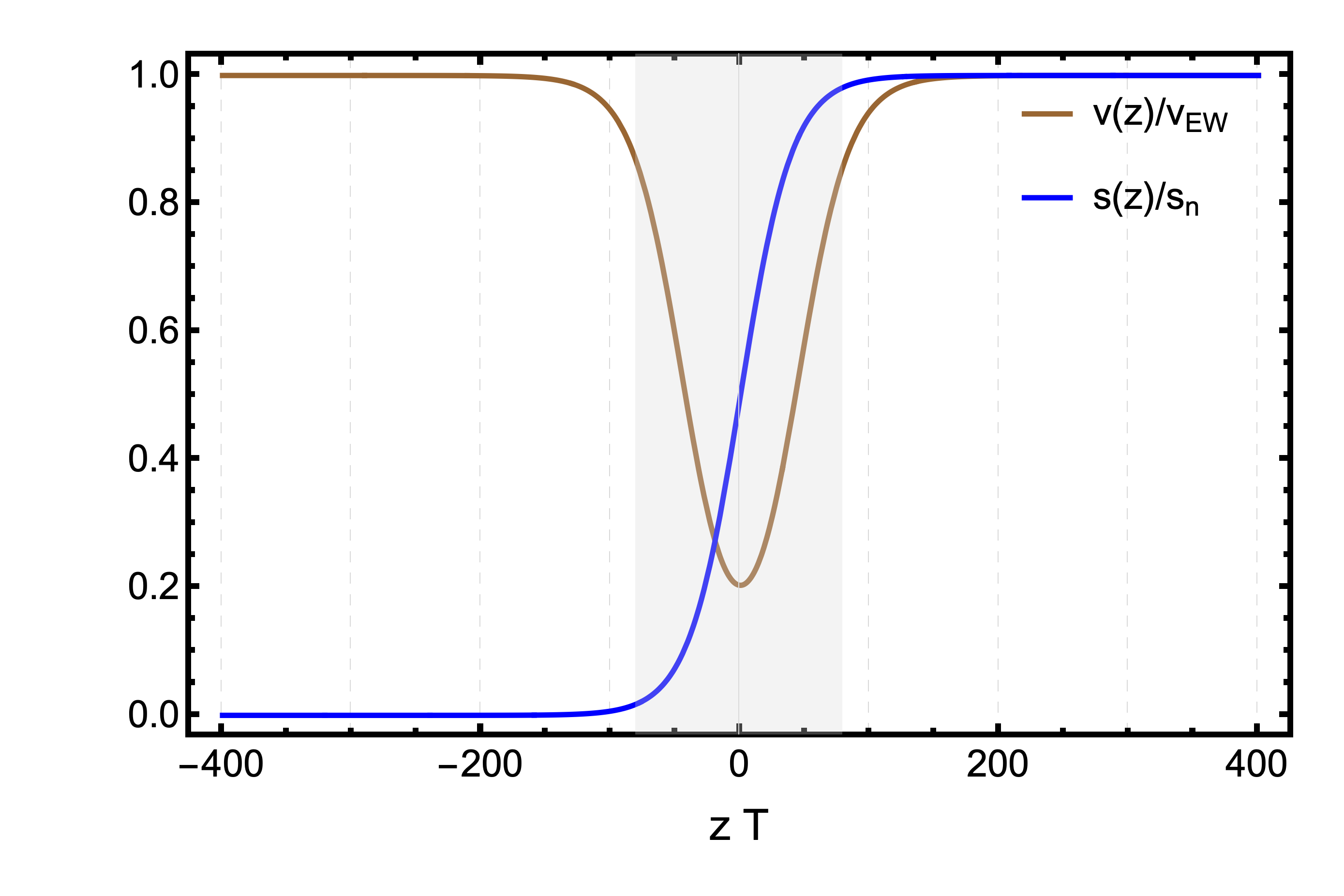}
        \label{fig:vevs}
    }
    \hfill
    \subfloat[Temperature dependence of baryon asymmetry produced with domain walls. The parameters are chosen as $h_0=20$ GeV, $\gamma_wv_w\approx0.4$, $s_n/\Lambda=0.2$ and $l_w=1 \text{ GeV}^{-1}$.]{
        \includegraphics[width=0.47\linewidth]{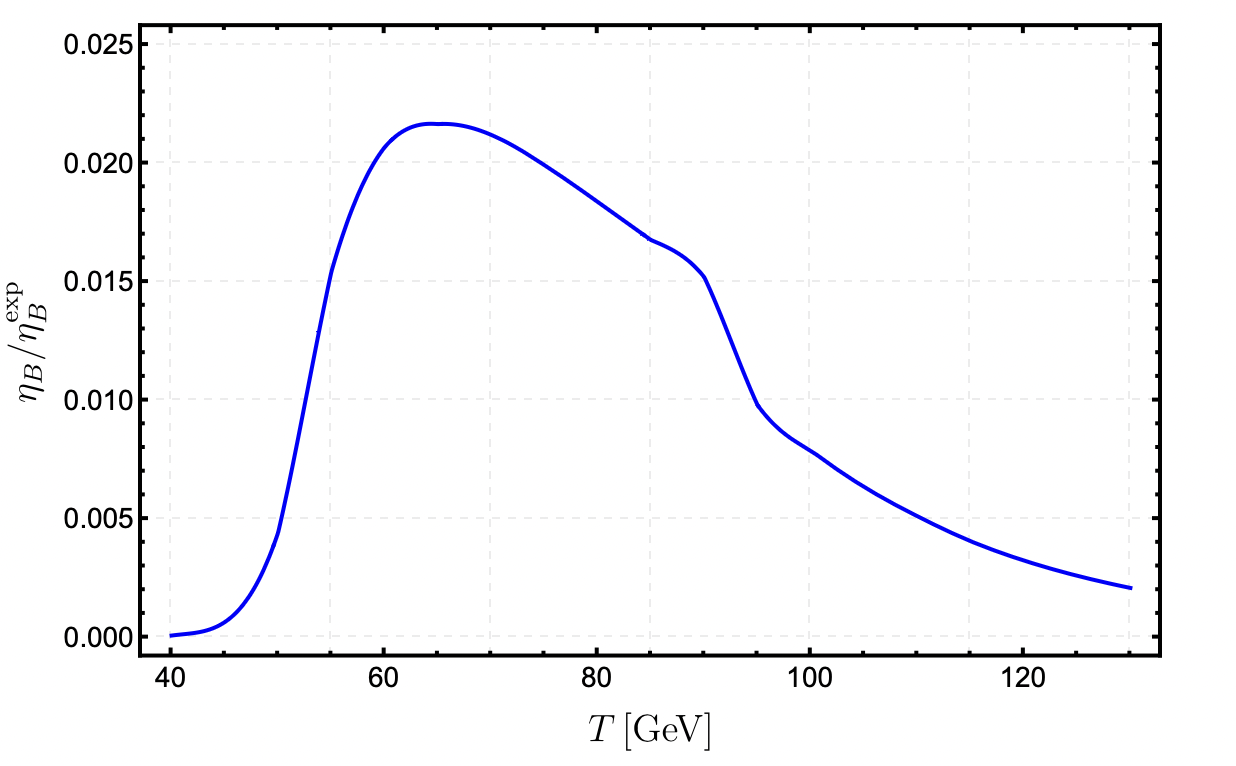}
        \label{fig:tdepdW}
    }
    \caption{Field profiles and temperature dependence of baryon asymmetry in the domain wall scenario.}
    \label{fig:combined}
\end{figure}\newline
The only requirement is that the length of the core must satisfy \cite{Azzola:2024pzq,Brandenberger:1996st}
\begin{equation}
    l_{core}>\left(\alpha_WT\right)^{-1},
\end{equation}
since the weak structure constant $\alpha_W\approx 0.033$, we see that $l_{core}\gtrsim 30/T$ and this places us well in the WKB regime discussed in the previous section.
The profile for $s(z)$ has been chosen as in Section \ref{Section: Toy model Sm with a singlet} and the two are sketched in Figure \ref{fig:vevs}. Other than that, the calculation proceeds in exactly the same way as described previously in this Appendix. The results\footnote{Note 
that the absolute value of baryon asymmetry is small in this case. This can be amplified by a different choice of $h_0$ or a different profile for the CPV source. However, this will not change the temperature dependence and we stick to the parameters chosen above.} are shown in Figure \ref{fig:tdepdW}. We can see that most of the baryon asymmetry will be generated mainly in the limited range of temperatures $T\in[50,100]\text{ GeV}$. Thus, even if the scaling regime lasts long for the purpose of BAU it is effectively very short making it strongly constrained. We are interested in computing the parameter
\bea
n_{\rm eff}= \frac{v}{\epsilon} \left[\frac{\l(\int \frac{dT}{T} \eta(T)\r)^2}{\int \frac{dT}{T} \eta^2(T)}\right],
\eea
which indeed turns out to be quite small,  $n_{\rm eff}\simeq0.58$.

\section{Different distributions of inhomogeneities}\label{Appendix: Different distributions}
In this Appendix, we present $\epsilon_{RMS}$ calculations for various baryon asymmetry distributions.
\subsection{$\delta$-function inhomogeneities}\label{Appendix: delta function inhomogeneities}
We start with the $\delta$-function like distribution, which has relevance for the models discussed in Section \ref{Section: Production via bubble collisions}.  The computation basically reproduces the results from Appendix A of \cite{Bagherian:2025puf}. The initial (before diffusion) density profile in a cubic region of  volume  $(NR)^3$ is given by:
\begin{equation}
    \rho(\vec{r},t=0)=\frac{1}{N^3}\sum_{n_x,n_y,n_z=-N/2}^{N/2}\delta^3(\vec{r}-R(n_x\hat{x}+n_y\hat{y}+n_z\hat{z})).
\end{equation}
The normalization is chosen such that the whole space has unit baryon asymmetry. With diffusion this evolves to
\begin{equation}
    \rho (\vec{r},t)=\frac{1}{(\sqrt{2\pi}Nd(t))^3}\sum_{\vec{n}}e^{-\frac{(\vec{r}-R\vec{n})^2}{2d(t)^2}}.
\end{equation}
We want to evaluate the RMS fluctuation, given by
\begin{equation}
    \epsilon_{RMS}^2=\frac{\langle \rho^2\rangle-\langle \rho \rangle^2}{\langle \rho \rangle^2}
\end{equation}
We first compute
\begin{equation}
\begin{split}
    \langle \rho \rangle=\frac{1}{(NR)^3}\int d^3 x \ \rho(\vec{r},t)&=\frac{1}{(\sqrt{2\pi}N^2d(t)R)^3} \sum_{\vec{n}}\int d^3x \ e^{-\frac{\vec{r}-R\vec{n}}{2d^2}}\\
    &=\frac{1}{(N^2R)^3}\sum_{\vec{n}}1\\
    &=\left(\frac{N+1}{N^2R}\right)^3.
\end{split}
\end{equation}
Computation of $\langle \rho^2\rangle$ is easier in momentum space
\begin{equation}
   \langle \rho^2\rangle=\frac{1}{(NR)^3} \int d^3x \rho^2(\vec{r},t)=\frac{1}{(NR)^3} \int \frac{d^3k}{(2\pi)^3}\lvert \tilde{\rho}(k,t)\rvert^2(\vec{r},t),
\end{equation}
where
\begin{equation}
    \tilde{\rho}(k,t)=\frac{1}{N^3}\sum_{\vec{n}}e^{i\vec{k}\cdot\vec{n}R}e^{-d^2k^2/2}.
\end{equation}
Then we get that the integral is
\begin{equation}
\begin{split}
    \langle \rho^2\rangle&=\frac{1}{(2\pi N^2R)^3}\sum_{\vec{n},\vec{n^\prime}}\int d^3 k\ e^{i\vec{k}\cdot (\vec{n}-\vec{n}^\prime)R}e^{-d^2k^2}\\
    &=\frac{1}{(2\sqrt{\pi} N^2Rd)^3}\sum_{\vec{n},\vec{n^\prime}}e^{-\frac{\lvert \vec{n}-\vec{n}^\prime\rvert^2R^2}{4d^2}}\\
    &=\left(\frac{1}{2\sqrt{\pi} N^2Rd}\sum_{n,n^\prime}e^{-\frac{( n-n^\prime)^2R^2}{4d^2}}\right)^3
\end{split}
\end{equation}
The sum is quite complicated, but we simplify it by summing over the difference $p=n-n^\prime$
\begin{equation}
    \sum_{n,n^\prime}e^{-\frac{( n-n^\prime)^2R^2}{4d^2}}=\sum_{p=-N}^{p=N}(N+1-\lvert p \rvert )e^{-R^2p^2/4d^2},
\end{equation}
where the term in the bracket is the degeneracy factor. This leads us to the final answer
\begin{equation}
    \langle \rho^2\rangle=\left(\frac{1}{2\sqrt{\pi} N^2Rd}\sum_{p=-N}^{p=N}(N+1-\lvert p \rvert )e^{-R^2p^2/4d^2}\right)^3.
\end{equation}
Combining all of this gives us the RMS fluctuation (after taking the limit $N\rightarrow\infty$)
\begin{equation}
    \epsilon_{RMS}^2=\left(\frac{R}{2\sqrt{\pi} d}\sum_{p=-\infty}^{\infty}e^{-R^2p^2/4d^2}\right)^3-1.
\end{equation}
Notice that the bound $\epsilon_{RMS}<0.3$ is saturated for roughly $R/d\sim 3$, which places a bound on the length scale of inhomogeneities
\begin{equation}
    R\lesssim3d.
\end{equation}
where $R$ is the comoving length scale of inhomogeneities.

\subsection{Spherical inhomogeneities}\label{spheres}
\begin{figure}[t!]
    \centering
    \subfloat[$\epsilon_{RMS}^2(R/R^{\rm min})$ for different values of $R^{\rm min}/d$.]{
        \includegraphics[width=0.46\textwidth]{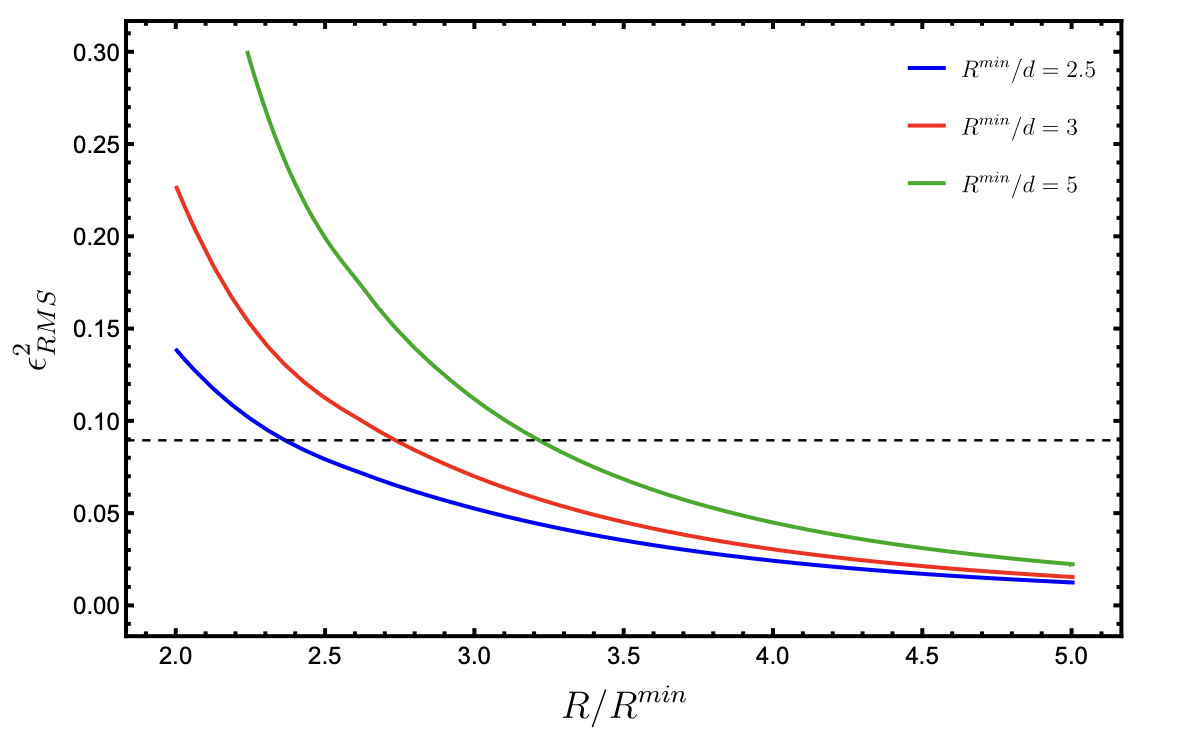}
        \label{fig:erms}
        }
    \hfill
    \subfloat[Allowed region in the parameter space represented by the shaded area.]{
        \includegraphics[width=0.46\textwidth]{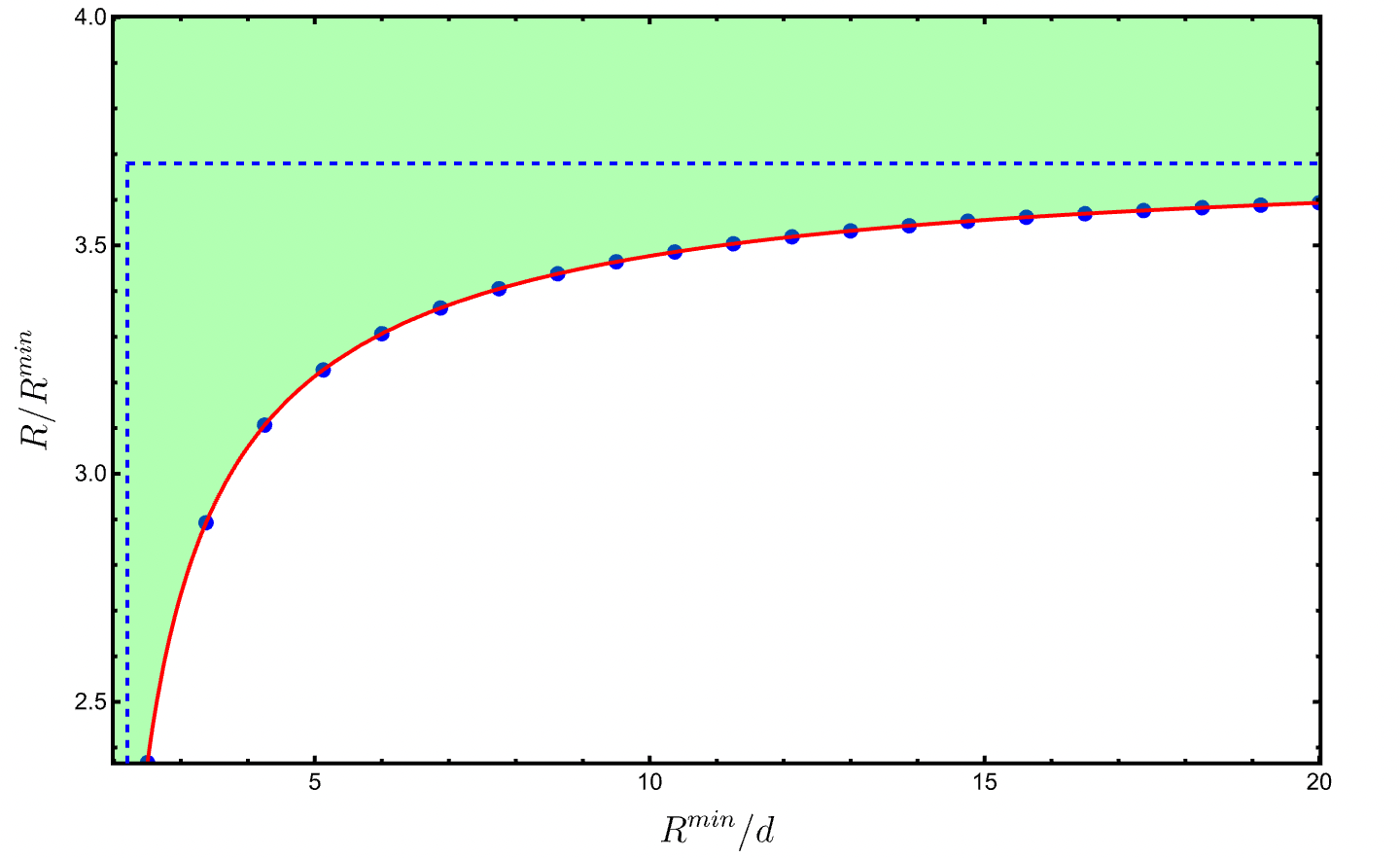}
        \label{fig:allowed}
    }
    \label{fig:comparison}
    \caption{}
\end{figure}

We follow a similar approach to that of the previous section and consider a density profile initially described by empty spheres of radius $R^{\rm min}$ placed on a lattice with spacing $R$. The density profile is given by
\begin{equation}
    \rho(r,t=0)=1-\sum_{\vec{n}}\theta(R^{\rm min}-\lvert \vec{x}-\vec{n}R\rvert).
\end{equation}
We perform the Fourier transform to obtain
\begin{equation}
    \tilde{\rho}(k,t)=\left[(2\pi)^3\delta^{(3)}(\vec{k})-\sum_{\vec{n}}e^{-i\vec{k}\cdot\vec{n}R}\int_{\lvert \vec{y}\rvert \leq R^{\rm min}} d^3 y\ e^{-i\vec{k}\cdot \vec{y}}\right] e^{-d^2k^2/2}.
\end{equation}
We Fourier transform it back into position space
\begin{equation}
\begin{split}
    \rho(r,t)&=1-\frac{1}{(2\pi)^3}\int d^3k\int_{\lvert \vec{y}\rvert<R^{\rm min}} d^3y\sum_{\vec{n}} e^{i\vec{k}\cdot(\vec{x}-\vec{y}-\vec{n}R)}e^{-d^2k^2/2}\\
    &=1-\frac{1}{(2\pi)^{3/2}d^3}\int_{\lvert \vec{y}\rvert<R^{\rm min}} d^3y\sum_{\vec{n}}e^{-(\vec{x}-\vec{n}R-\vec{y})^2/(2d^2)}
\end{split}
\end{equation}
Focusing on the last integral ($\vec{p}=\vec{x}-\vec{n}R$, $p=\lvert \vec{p}\rvert$, $y=\lvert \vec{y}\rvert$)
\begin{equation}
\begin{split}
    \int_{\lvert \vec{y}\rvert<R^{\rm min}} d^3y\sum_{\vec{n}}e^{-(\vec{x}-\vec{n}R-\vec{y})^2/(2d^2)}&=2\pi\int_0^{R^{\rm min}}dy\ y^2\int_{-1}^1d (\cos \theta)e^{-(p^2+y^2-2py\cos\theta)/(2d^2)}\\
    &=\frac{4\pi d^2}{p} \int_0^{R^{\rm min}} dy \ y\ e^{-(p^2+y^2)/(2d^2)}\sinh \left(\frac{py}{d^2}\right)
\end{split}
\end{equation}
Performing this integral explicitly leads us to the final result
\begin{equation}
\begin{split}
    \rho(r,t)=&1-\sum_{\vec{n}}\bigg[\frac{1}{2}\left(\text{erf}\left(\frac{\lvert \vec{x}-\vec{n}R\rvert+R^{\rm min}}{\sqrt{2}d}\right)-\text{erf}\left(\frac{\lvert \vec{x}-\vec{n}R\rvert-R^{\rm min}}{\sqrt{2}d}\right)\right)\\
    &+\frac{1}{\sqrt{2\pi}}\frac{d}{\lvert \vec{x}-\vec{n}R\rvert}\left(e^{-\left(\frac{\lvert \vec{x}-\vec{n}R\rvert+R^{\rm min}}{\sqrt{2}d}\right)^2}-e^{-\left(\frac{\lvert \vec{x}-\vec{n}R\rvert-R^{\rm min}}{\sqrt{2}d}\right)^2}\right)\bigg].
\end{split}
\end{equation}
Averaging over space
\begin{equation}
    \langle \rho \rangle =1-\frac{4\pi (R^{\rm min})^3}{3R^3}.
\end{equation}
The average $\langle \rho^2\rangle$ is again easier to compute in momentum space
\begin{equation}
\begin{split}
    \langle \rho^2\rangle&=\frac{1}{V}\int d^3x \ \rho(x,t) ^2\\
    &=1-\frac{8\pi (R^{\rm min})^3}{3R^3}+\frac{1}{V}\sum_{\vec{n},\vec{m}}\int_{y_{1,2}<R^{\rm min}}d^3y_1 \ d^3y_2\int \frac{d^3q} {(2\pi)^3}e^{-d^2q^2}e^{-i\vec{q}\cdot[(\vec{y}_1-\vec{y}_2)+\vec{\Delta}R]}\\
    &=1-\frac{8\pi (R^{\rm min})^3}{3R^3}+\frac{1}{V(2\sqrt{\pi}d)^3}\sum_{\vec{n},\vec{m}}\int_{y_{1,2}<R^{\rm min}}d^3y_1 \ d^3y_2\ e^{-\lvert(\vec{y}_1-\vec{y}_2)+\vec{\Delta}R\rvert^2/(4d^2)}
\end{split}
\end{equation}
where $\vec{\Delta}= \vec{n}-\vec{m}$. Computation of the integral is a bit involved, to perform it we do the substitution $\vec{s}=\vec{y}_1-\vec{y}_2$ and $\vec{u}=\vec{y}_2$, then
\begin{equation}
    \int_{y_{1,2}<R^{\rm min}}d^3y_1 \ d^3y_2\ e^{-\lvert(\vec{y}_1-\vec{y}_2)+\vec{\Delta}R\rvert^2/(4d^2)}=\int d^3s \ e^{-\lvert\vec{\Delta}R+\vec{s}\rvert/(4d^2)}\int_{\vec{u}\in B_{R^{\rm min}},\ \vec{s}+\vec{u} \in B_{R^{\rm min}}} d^3 u.
\end{equation}
Since we are integrating over two spheres, the integral over $\vec{u}$ corresponds to the volume enclosed by the intersection of two spheres, where $\vec{s}$ is the vector which points from one centre to the other. Hence, the second integral is just the volume of the two spherical caps
\begin{equation}
\int_{\vec{u}\in B_{R^{\rm min}},\ \vec{s}+\vec{u} \in B_{R^{\rm min}}} d^3 u\equiv F(s)=\frac{\pi}{12}s(4R^{\rm min}+s)(2R^{\rm min}-s)^2.
\end{equation}
Performing the integral over the angular part of $\vec{s}$
\begin{equation}
\frac{1}{\sqrt{\pi}d R^4}\sum_{\vec{n},\vec{m}}\frac{1}{\Delta}\int _0^{2R^{\rm min}} F(s)e^{-(\Delta^2R^2+s^2)/(4d^2)}\sinh \left(\frac{\Delta R s}{2d^2}\right).
\end{equation}
Since our sum only depends on the difference $\vec{\Delta}=\vec{n}-\vec{m}$, we can rewrite this as a sum over $\vec{\Delta}$ times the appropriate degeneracy factor ($f(\Delta)$ being our term in the sum)
\begin{equation}   \sum_{\vec{n},\vec{m}}f(\Delta)=\sum_{\vec{\Delta}}(N+1-\lvert\Delta_x\rvert)(N+1-\lvert\Delta_y\rvert)(N+1-\lvert\Delta_z\rvert)f(\Delta)\approx N^3\sum_{\vec{\Delta}}f(\Delta),
\end{equation}
where we use large $N$ approximation\footnote{This approximation certainly works even for large $\Delta_i$, because of the Gaussian suppression of $f(\Delta)$.}. This leads to the final result in the large $N$ limit
\begin{equation}
\begin{split}
    \langle \rho^2\rangle&=1-\frac{8\pi (R^{\rm min})^3}{3R^3}+\frac{1}{\sqrt{\pi}d R^4}\sum_{\vec{\Delta}}\frac{1}{\Delta}\int _0^{2R^{\rm min}} F(s)e^{-(\Delta^2R^2+s^2)/(4d^2)}\sinh \left(\frac{\Delta R s}{2d^2}\right),
\end{split}
\end{equation}
where $\Delta=\lvert \vec{n}-\vec{m}\rvert$. Focusing on the last term, since it depends only on the length of the vector $\vec{\Delta}$, we can approximate it with the following
\begin{equation}
    \frac{1}{\sqrt{\pi}dR^4}\sum_{\vec{\Delta}}(\dots)\approx\frac{1}{\sqrt{\pi}dR^4}\left[(\Delta\rightarrow0)+\sum_{\Delta=1}^{\sqrt{3}\Delta_0}g(\Delta)(\dots)+\int_{\sqrt{3}\Delta_0}^{\infty} d\Delta (4\pi \Delta^2) (\dots)\right],
\end{equation}
where $g(\Delta)$ is the degeneracy of vectors with the same length and $\Delta_0$ is some cut-off from which we use the continuum limit. Results are plotted in Figures \ref{fig:erms} and \ref{fig:allowed}. 
The curve on Figure \ref{fig:allowed} is well-fitted by
\begin{equation}
    y(x)=3.68-24.56 \exp\left[-3.13 (x-1.79)^{0.2}\right].
\end{equation}
This is an involved curve, but it is instructed to look at some limiting cases. For large $x$'s (large $R^{\rm min}/d$), the exponential can be ignored and we find the bound
\begin{equation}
    \frac{R}{R^{\rm min}}>3.68.
\end{equation}
There is another bound which we can extract, for small $R^{\rm min}/d$, $\epsilon_{RMS}^2(R/R^{\rm min})<0.09$ for all $R/R^{\rm min}>2$. Using this we get the following bound
\begin{equation}
    \frac{R^{\rm min}}{d(T_{BBN})}<2.2.
\end{equation}
\subsection{Domain walls: annihilation}\label{Appendix: DW annhiliation}
Once we add the bias to the model, favoring one vacuum over the other, domain walls begin to move and hence there will be a production of baryons in the regions where the non-favourable vacuum was. This will leave some inhomogeneity of baryons, which we model with a "2D chess board" density profile
\begin{equation}
    \rho(\vec{x},t)=\frac{1}{2}\left[1+\text{sgn}\left(\cos\left(\frac{\pi x}{D}\right)\cos\left(\frac{\pi y}{D}\right)\right)\right].
\end{equation}
The difference between this model and the other models is in the fact that we know that $D$ is of horizon size. Going to momentum space we find that
\begin{equation}
    \tilde{\rho}(\vec{k},0)=\frac{1}{2}\left[(2\pi)^2\delta^{(2)}(\vec{k})+\int d^2x \ e^{-i\vec{k}\cdot\vec{x}}\text{sgn}\left(\cos\left(\frac{\pi x}{D}\right)\cos\left(\frac{\pi y}{D}\right)\right)\right].
\end{equation}
Of course, it is true that $\text{sgn}(pqr)=\text{sgn}(p)\text{sgn}(q)$, hence we have two integrals of the form
\begin{equation}
    \int_{-\infty}^\infty dx \ e^{-ikx} \text{sgn}\left(\cos\left(\frac{\pi x}{D}\right)\right)=4\sum_{m=0}^\infty\frac{(-1)^m}{2m+1} \left[\delta\left(k-\frac{(2m+1)\pi}{D}\right)+\delta\left(k+\frac{(2m+1)\pi}{D}\right)\right],
\end{equation}
where we used the fact that $\text{sgn}(\cos(\theta))$ can be written in a Fourier series
\begin{equation}
    \text{sgn}(\cos(\theta))=\frac{4}{\pi}\sum_{m=0}^\infty\frac{(-1)^m}{2m+1}\cos((2m+1)\theta).
\end{equation}
We obtain that
\begin{equation}
    \tilde{\rho}(\vec{k},t=0)=\frac{1}{2}\left\{(2\pi)^2\delta^{(3)}(\vec{k})+\prod_{k_i}4\sum_{m=0}^\infty\frac{(-1)^m}{2m+1}\left[\delta\left(k_i-\frac{(2m+1)\pi}{D}\right)+\delta\left(k_i+\frac{(2m+1)\pi}{D}\right)\right]\right\},
\end{equation}
where $k_i\in\{k_x,k_y\}$. It is easy to see that going back to position space we get
\begin{equation}
\begin{split}
    \rho(\vec{x},t)&=\frac{1}{(2\pi)^2}\int d^2k\ e^{i\vec{k}\cdot\vec{x}} e^{-d^2k^2/2}\tilde{\rho}(k,0)\\
    &=\frac{1}{2}\left[1+\prod_{x_i\in\{x,y\}}\frac{4}{\pi}\sum_{m=0}^\infty\frac{(-1)^m}{2m+1}e^{-\frac{d^2(2m+1)^2\pi^2}{2D^2}}\cos \left(\frac{(2m+1)\pi}{D}x_i\right)\right]
\end{split}
\end{equation}
Averaging this over space we get $\langle \rho \rangle=1/2$. As usual $\langle \rho^2\rangle$ is easier to compute in momentum space
\begin{equation}
\begin{split}
    \langle\rho^2\rangle=\frac{1}{4}+\frac{1}{4V}\int\frac{d^2k}{(2\pi)^2}\ e^{-d^2k^2}&\prod_{k_i}4\sum_{m=0}^\infty\frac{(-1)^m}{2m+1}\left[\delta\left(k_i-\frac{(2m+1)\pi}{D}\right)+\delta\left(k_i+\frac{(2m+1)\pi}{D}\right)\right]\\
    \times&\prod_{k_i}4\sum_{n=0}^\infty\frac{(-1)^n}{2n+1}\left[\delta\left(-k_i-\frac{(2n+1)\pi}{D}\right)+\delta\left(-k_i+\frac{(2n+1)\pi}{D}\right)\right].
    \end{split}
\end{equation}
Again, we have two terms multiplying of the form
\begin{equation}
\begin{split}
    \frac{8}{\pi}\sum_{m,n=0}^\infty &\frac{(-1)^{m+n}}{(2m+1)(2n+1)}\int_{-\infty}^\infty dk \ e^{-d^2k^2} \\
   \times \bigg[&\delta\left(k-\frac{(2m+1)\pi}{D}\right)\delta\left(-k-\frac{(2n+1)\pi}{D}\right)+\delta\left(k-\frac{(2m+1)\pi}{D}\right)\delta\left(-k+\frac{(2n+1)\pi}{D}\right)\\
    +&\delta\left(k+\frac{(2m+1)\pi}{D}\right)\delta\left(-k-\frac{(2n+1)\pi}{D}\right)+\delta\left(k+\frac{(2m+1)\pi}{D}\right)\delta\left(-k+\frac{(2n+1)\pi}{D}\right)\bigg].\\
\end{split}
\end{equation}
It is easy to see that the first and last term survive when $m+n+1=0$ (which cannot happen since $m,n>0$), whereas second and third are non-zero when $m=n$. Finally using the $\delta$-function to compute the integral over $k$ we get
\begin{equation}
    \frac{8}{\pi}\sum_{m,n=0}^\infty (\dots)=\frac{16\delta(0)}{\pi}\sum_{m=0}^\infty \frac{1}{(2m+1)^2}e^{-d^2(2m+1)^2\pi^2/D^2}.
\end{equation}
This gives us
\begin{equation}
    \langle \rho^2\rangle=\frac{1}{4}+\frac{1}{4}\frac{\delta^{(2)}(0)}{V}\left(\frac{16}{\pi}\sum_{m=0}^\infty \frac{1}{(2m+1)^2}e^{-d^2(2m+1)^2\pi^2/D^2}\right)^2.
\end{equation}
It is well-known that the momentum space $\delta$-function $\lim_{\vec{k}\rightarrow 0} \delta^{(2)}(\vec{k})=V/(2\pi)^2$, which leads us to the final result
\begin{equation}
     \langle \rho^2\rangle=\frac{1}{4}+\frac{1}{4}\left(\frac{8}{\pi^2}\sum_{m=0}^\infty \frac{1}{(2m+1)^2}e^{-d^2(2m+1)^2\pi^2/D^2}\right)^2,
\end{equation}
which gives the RMS fluctuation to be simply
\begin{equation}
    \epsilon_{RMS}^2=\left(\frac{8}{\pi^2}\sum_{m=0}^\infty \frac{1}{(2m+1)^2}e^{-d^2(2m+1)^2\pi^2/D^2}\right)^2.
\end{equation}
Imposing the deuterium bounds, we get
\begin{equation}
    \frac{d(T_{BBN})}{D}>0.317.
\end{equation}
Since these are formed via Kibble mechanism, $D$ is given by the comoving horizon size $D\sim 1/(aH)$ at the temperature of domain-wall annihilation
\begin{equation}
    D\sim\frac{1}{(aH)}\bigg\rvert_{T=T_{ann}}=\frac{3}{\pi}\left(\frac{10}{g_*(T)}\right)^{1/2}\left(\frac{g_{\ast S}(T)}{g_{\ast S}(T_0)}\right)^{1/3}\frac{M_{Pl}}{T_0T_{ann}},
\end{equation}
this leads to a bound on the temperature of domain-wall annihilation (using the diffusion length of protons)
\begin{equation}
    T_{ann}\gtrsim1.7\text{ TeV}.
\end{equation}

\bibliographystyle{JHEP}
\bibliography{biblio}

@book{Mukhanov:2005sc,
    author = "Mukhanov, V.",
    title = "{Physical Foundations of Cosmology}",
    doi = "10.1017/CBO9780511790553",
    isbn = "978-0-521-56398-7",
    publisher = "Cambridge University Press",
    address = "Oxford",
    year = "2005"
}

@article{Gouttenoire:2023naa,
    author = "Gouttenoire, Yann and Volansky, Tomer",
    title = "{Primordial black holes from supercooled phase transitions}",
    eprint = "2305.04942",
    archivePrefix = "arXiv",
    primaryClass = "hep-ph",
    doi = "10.1103/PhysRevD.110.043514",
    journal = "Phys. Rev. D",
    volume = "110",
    number = "4",
    pages = "043514",
    year = "2024"
}

@article{Liu:2021svg,
    author = "Liu, Jing and Bian, Ligong and Cai, Rong-Gen and Guo, Zong-Kuan and Wang, Shao-Jiang",
    title = "{Primordial black hole production during first-order phase transitions}",
    eprint = "2106.05637",
    archivePrefix = "arXiv",
    primaryClass = "astro-ph.CO",
    doi = "10.1103/PhysRevD.105.L021303",
    journal = "Phys. Rev. D",
    volume = "105",
    number = "2",
    pages = "L021303",
    year = "2022"
}

@article{Bagherian:2025puf,
    author = "Bagherian, Hengameh and Ekhterachian, Majid and Stelzl, Stefan",
    title = "{The Bearable Inhomogeneity of the Baryon Asymmetry}",
    eprint = "2505.15904",
    archivePrefix = "arXiv",
    primaryClass = "hep-ph",
    month = "5",
    year = "2025"
}

@article{Applegate:1987hm,
    author = "Applegate, J. H. and Hogan, C. J. and Scherrer, R. J.",
    title = "{Cosmological Baryon Diffusion and Nucleosynthesis}",
    doi = "10.1103/PhysRevD.35.1151",
    journal = "Phys. Rev. D",
    volume = "35",
    pages = "1151--1160",
    year = "1987"
}

@article{Enqvist:1991xw,
    author = "Enqvist, K. and Ignatius, J. and Kajantie, K. and Rummukainen, K.",
    title = "{Nucleation and bubble growth in a first order cosmological electroweak phase transition}",
    reportNumber = "HU-TFT-91-35",
    doi = "10.1103/PhysRevD.45.3415",
    journal = "Phys. Rev. D",
    volume = "45",
    pages = "3415--3428",
    year = "1992"
}

@article{Riotto:1999yt,
    author = "Riotto, Antonio and Trodden, Mark",
    title = "{Recent progress in baryogenesis}",
    eprint = "hep-ph/9901362",
    archivePrefix = "arXiv",
    reportNumber = "CERN-TH-99-04, CWRU-P6-99",
    doi = "10.1146/annurev.nucl.49.1.35",
    journal = "Ann. Rev. Nucl. Part. Sci.",
    volume = "49",
    pages = "35--75",
    year = "1999"
}

@article{White:2016nbo,
    author = "White, Graham Albert",
    title = "{A Pedagogical Introduction to Electroweak Baryogenesis}",
    doi = "10.1088/978-1-6817-4457-5",
    month = "11",
    year = "2016"
}

@article{Garbrecht:2018mrp,
    author = {Garbrecht, Bj{\"o}rn},
    title = "{Why is there more matter than antimatter? Calculational methods for leptogenesis and electroweak baryogenesis}",
    eprint = "1812.02651",
    archivePrefix = "arXiv",
    primaryClass = "hep-ph",
    reportNumber = "TUM-HEP-1177-18",
    doi = "10.1016/j.ppnp.2019.103727",
    journal = "Prog. Part. Nucl. Phys.",
    volume = "110",
    pages = "103727",
    year = "2020"
}

@article{Dine:2003ax,
    author = "Dine, Michael and Kusenko, Alexander",
    title = "{The Origin of the matter - antimatter asymmetry}",
    eprint = "hep-ph/0303065",
    archivePrefix = "arXiv",
    reportNumber = "SCIPP-2003-08, UCLA-03-TEP-08",
    doi = "10.1103/RevModPhys.76.1",
    journal = "Rev. Mod. Phys.",
    volume = "76",
    pages = "1",
    year = "2003"
}

@article{Trodden:1998ym,
    author = "Trodden, Mark",
    title = "{Electroweak baryogenesis}",
    eprint = "hep-ph/9803479",
    archivePrefix = "arXiv",
    reportNumber = "CWRU-P6-98",
    doi = "10.1103/RevModPhys.71.1463",
    journal = "Rev. Mod. Phys.",
    volume = "71",
    pages = "1463--1500",
    year = "1999"
}

@article{Katz:2016adq,
    author = "Katz, Andrey and Riotto, Antonio",
    title = "{Baryogenesis and Gravitational Waves from Runaway Bubble Collisions}",
    eprint = "1608.00583",
    archivePrefix = "arXiv",
    primaryClass = "hep-ph",
    reportNumber = "CERN-TH-2016-173",
    doi = "10.1088/1475-7516/2016/11/011",
    journal = "JCAP",
    volume = "11",
    pages = "011",
    year = "2016"
}

@article{Cataldi:2024pgt,
    author = "Cataldi, Martina and Shakya, Bibhushan",
    title = "{Leptogenesis via bubble collisions}",
    eprint = "2407.16747",
    archivePrefix = "arXiv",
    primaryClass = "hep-ph",
    reportNumber = "DESY-24-110",
    doi = "10.1088/1475-7516/2024/11/047",
    journal = "JCAP",
    volume = "11",
    pages = "047",
    year = "2024"
}

@article{Azatov:2022tii,
    author = "Azatov, Aleksandr and Barni, Giulio and Chakraborty, Sabyasachi and Vanvlasselaer, Miguel and Yin, Wen",
    title = "{Ultra-relativistic bubbles from the simplest Higgs portal and their cosmological consequences}",
    eprint = "2207.02230",
    archivePrefix = "arXiv",
    primaryClass = "hep-ph",
    reportNumber = "SISSA 12/2022/FISI TU-1157",
    doi = "10.1007/JHEP10(2022)017",
    journal = "JHEP",
    volume = "10",
    pages = "017",
    year = "2022"
}

@article{Azzola:2024pzq,
    author = "Azzola, Jacopo and Matsedonskyi, Oleksii and Weiler, Andreas",
    title = "{Minimal electroweak baryogenesis via domain walls}",
    eprint = "2412.10495",
    archivePrefix = "arXiv",
    primaryClass = "hep-ph",
    doi = "10.1007/JHEP04(2025)103",
    journal = "JHEP",
    volume = "04",
    pages = "103",
    year = "2025"
}

@article{Falkowski:2012fb,
    author = "Falkowski, Adam and No, Jose M.",
    title = "{Non-thermal Dark Matter Production from the Electroweak Phase Transition: Multi-TeV WIMPs and 'Baby-Zillas'}",
    eprint = "1211.5615",
    archivePrefix = "arXiv",
    primaryClass = "hep-ph",
    reportNumber = "ULB-TH-12-18, LPT-12-113",
    doi = "10.1007/JHEP02(2013)034",
    journal = "JHEP",
    volume = "02",
    pages = "034",
    year = "2013"
}

@article{Cataldi:2025nac,
    author = {Cataldi, Martina and M{\"u}{\"u}rsepp, Kristjan and Vanvlasselaer, Miguel},
    title = "{When bubbles collide}",
    eprint = "2506.12123",
    archivePrefix = "arXiv",
    primaryClass = "hep-ph",
    reportNumber = "DESY-25-082",
    month = "6",
    year = "2025"
}

@article{Azatov:2021irb,
    author = "Azatov, Aleksandr and Vanvlasselaer, Miguel and Yin, Wen",
    title = "{Baryogenesis via relativistic bubble walls}",
    eprint = "2106.14913",
    archivePrefix = "arXiv",
    primaryClass = "hep-ph",
    reportNumber = "SISSA 13/2021/FISI TU-1127",
    doi = "10.1007/JHEP10(2021)043",
    journal = "JHEP",
    volume = "10",
    pages = "043",
    year = "2021"
}

@article{Shakya:2023kjf,
    author = "Shakya, Bibhushan",
    title = "{Aspects of particle production from bubble dynamics at a first order phase transition}",
    eprint = "2308.16224",
    archivePrefix = "arXiv",
    primaryClass = "hep-ph",
    reportNumber = "DESY 23-125",
    doi = "10.1103/PhysRevD.111.023521",
    journal = "Phys. Rev. D",
    volume = "111",
    number = "2",
    pages = "023521",
    year = "2025"
}

@article{Mansour:2023fwj,
    author = "Mansour, Henda and Shakya, Bibhushan",
    title = "{Particle production from phase transition bubbles}",
    eprint = "2308.13070",
    archivePrefix = "arXiv",
    primaryClass = "hep-ph",
    reportNumber = "DESY-23-121, TTP23-033, P3H-23-056",
    doi = "10.1103/PhysRevD.111.023520",
    journal = "Phys. Rev. D",
    volume = "111",
    number = "2",
    pages = "023520",
    year = "2025"
}

@article{Mariotti:2024eoh,
    author = {Mariotti, Alberto and Nagels, Xander and Rase, A{\"a}ron and Vanvlasselaer, Miguel},
    title = "{DW-genesis: baryon number from domain wall network collapse}",
    eprint = "2411.13494",
    archivePrefix = "arXiv",
    primaryClass = "hep-ph",
    doi = "10.1007/JHEP03(2025)199",
    journal = "JHEP",
    volume = "03",
    pages = "199",
    year = "2025"
}

@article{Daido:2015gqa,
    author = "Daido, Ryuji and Kitajima, Naoya and Takahashi, Fuminobu",
    title = "{Axion domain wall baryogenesis}",
    eprint = "1504.07917",
    archivePrefix = "arXiv",
    primaryClass = "hep-ph",
    reportNumber = "TU-993, IPMU15-0059",
    doi = "10.1088/1475-7516/2015/07/046",
    journal = "JCAP",
    volume = "07",
    pages = "046",
    year = "2015"
}

@article{Avelino:2005kn,
    author = "Avelino, P. P. and Martins, C. J. A. P. and Oliveira, J. C. R. E.",
    title = "{One-scale model for domain wall network evolution}",
    eprint = "hep-ph/0507272",
    archivePrefix = "arXiv",
    doi = "10.1103/PhysRevD.72.083506",
    journal = "Phys. Rev. D",
    volume = "72",
    pages = "083506",
    year = "2005"
}

@article{Kawano:1989mw,
    author = "Kawano, Lawrence",
    title = "{Evolution of Domain Walls in the Early Universe}",
    reportNumber = "FERMILAB-THESIS-1989-09, FERMILAB-PUB-89-208-A",
    doi = "10.1103/PhysRevD.41.1013",
    journal = "Phys. Rev. D",
    volume = "41",
    pages = "1013",
    year = "1990"
}

@article{Press:1989yh,
    author = "Press, William H. and Ryden, Barbara S. and Spergel, David N.",
    title = "{Dynamical Evolution of Domain Walls in an Expanding Universe}",
    reportNumber = "NSF-ITP-89-51, CFA-1870",
    doi = "10.1086/168151",
    journal = "Astrophys. J.",
    volume = "347",
    pages = "590--604",
    year = "1989"
}

@article{Dankovsky:2024zvs,
    author = "Dankovsky, I. and Babichev, E. and Gorbunov, D. and Ramazanov, S. and Vikman, A.",
    title = "{Revisiting evolution of domain walls and their gravitational radiation with CosmoLattice}",
    eprint = "2406.17053",
    archivePrefix = "arXiv",
    primaryClass = "astro-ph.CO",
    doi = "10.1088/1475-7516/2024/09/047",
    journal = "JCAP",
    volume = "09",
    pages = "047",
    year = "2024"
}

@article{Blasi:2025tmn,
    author = {Blasi, Simone and Mariotti, Alberto and Rase, A{\"a}ron and Vanvlasselaer, Miguel},
    title = "{Domain walls in the scaling regime: Equal Time Correlator and Gravitational Waves}",
    eprint = "2511.16649",
    archivePrefix = "arXiv",
    primaryClass = "hep-ph",
    month = "11",
    year = "2025"
}

@article{Kibble:1976sj,
    author = "Kibble, T. W. B.",
    title = "{Topology of Cosmic Domains and Strings}",
    reportNumber = "ICTP/75/5",
    doi = "10.1088/0305-4470/9/8/029",
    journal = "J. Phys. A",
    volume = "9",
    pages = "1387--1398",
    year = "1976"
}

@article{Leite:2011sc,
    author = "Leite, A. M. M. and Martins, C. J. A. P.",
    title = "{Scaling Properties of Domain Wall Networks}",
    eprint = "1110.3486",
    archivePrefix = "arXiv",
    primaryClass = "hep-ph",
    doi = "10.1103/PhysRevD.84.103523",
    journal = "Phys. Rev. D",
    volume = "84",
    pages = "103523",
    year = "2011"
}

@article{Schroder:2024gsi,
    author = {Schr{\"o}der, Tobias and Brandenberger, Robert},
    title = "{Embedded domain walls and electroweak baryogenesis}",
    eprint = "2404.13035",
    archivePrefix = "arXiv",
    primaryClass = "hep-ph",
    doi = "10.1103/PhysRevD.110.043516",
    journal = "Phys. Rev. D",
    volume = "110",
    number = "4",
    pages = "043516",
    year = "2024"
}

@article{Brandenberger:1996st,
  title = {Local and nonlocal defect-mediated electroweak baryogenesis},
  author = {Brandenberger, Robert and Davis, Anne-Christine and Prokopec, Tomislav and Trodden, Mark},
  journal = {Phys. Rev. D},
  volume = {53},
  issue = {8},
  pages = {4257--4266},
  numpages = {0},
  year = {1996},
  month = {Apr},
  publisher = {American Physical Society},
  doi = {10.1103/PhysRevD.53.4257},
  url = {https://link.aps.org/doi/10.1103/PhysRevD.53.4257}
}

@article{Espinosa:2011eu,
    author = "Espinosa, Jose R. and Gripaios, Ben and Konstandin, Thomas and Riva, Francesco",
    title = "{Electroweak Baryogenesis in Non-minimal Composite Higgs Models}",
    eprint = "1110.2876",
    archivePrefix = "arXiv",
    primaryClass = "hep-ph",
    doi = "10.1088/1475-7516/2012/01/012",
    journal = "JCAP",
    volume = "01",
    pages = "012",
    year = "2012"
}

@article{Bodeker:2004ws,
    author = "Bodeker, Dietrich and Fromme, Lars and Huber, Stephan J. and Seniuch, Michael",
    title = "{The Baryon asymmetry in the standard model with a low cut-off}",
    eprint = "hep-ph/0412366",
    archivePrefix = "arXiv",
    reportNumber = "BI-TP-2004-41, CERN-PH-TH-2004-258",
    doi = "10.1088/1126-6708/2005/02/026",
    journal = "JHEP",
    volume = "02",
    pages = "026",
    year = "2005"
}

@article{Barni:2025ifb,
    author = "Barni, Giulio",
    title = "{Electroweak Baryogenesis with BARYONET: a self-contained review of the WKB approach}",
    eprint = "2510.21915",
    archivePrefix = "arXiv",
    primaryClass = "hep-ph",
    month = "10",
    year = "2025"
}

@article{Cline:2020jre,
    author = "Cline, James M. and Kainulainen, Kimmo",
    title = "{Electroweak baryogenesis at high bubble wall velocities}",
    eprint = "2001.00568",
    archivePrefix = "arXiv",
    primaryClass = "hep-ph",
    reportNumber = "CERN-TH-2019-227",
    doi = "10.1103/PhysRevD.101.063525",
    journal = "Phys. Rev. D",
    volume = "101",
    number = "6",
    pages = "063525",
    year = "2020"
}

@article{Cline:2000nw,
    author = "Cline, James M. and Joyce, Michael and Kainulainen, Kimmo",
    title = "{Supersymmetric electroweak baryogenesis}",
    eprint = "hep-ph/0006119",
    archivePrefix = "arXiv",
    reportNumber = "MCGILL-00-15, NORDITA-2000-38-HE, LPT-ORSAY-00-46",
    doi = "10.1088/1126-6708/2000/07/018",
    journal = "JHEP",
    volume = "07",
    pages = "018",
    year = "2000"
}

@article{Kainulainen:2024qpm,
    author = "Kainulainen, Kimmo and Venkatesan, Niyati",
    title = "{Systematic moment expansion for electroweak baryogenesis}",
    eprint = "2407.13639",
    archivePrefix = "arXiv",
    primaryClass = "hep-ph",
    doi = "10.1088/1475-7516/2024/08/058",
    journal = "JCAP",
    volume = "08",
    pages = "058",
    year = "2024"
}

@article{Cline:2017qpe,
    author = "Cline, James M. and Kainulainen, Kimmo and Tucker-Smith, David",
    title = "{Electroweak baryogenesis from a dark sector}",
    eprint = "1702.08909",
    archivePrefix = "arXiv",
    primaryClass = "hep-ph",
    reportNumber = "CERN-TH-2017-050",
    doi = "10.1103/PhysRevD.95.115006",
    journal = "Phys. Rev. D",
    volume = "95",
    number = "11",
    pages = "115006",
    year = "2017"
}

@article{vandeVis:2025efm,
    author = "van de Vis, Jorinde and de Vries, Jordy and Postma, Marieke",
    title = "{Bubble Trouble: a Review on Electroweak Baryogenesis}",
    eprint = "2508.09989",
    archivePrefix = "arXiv",
    primaryClass = "hep-ph",
    reportNumber = "CERN-TH-2025-161, Nikhef 2025-012",
    month = "8",
    year = "2025"
}

@article{Azatov:2020ufh,
    author = "Azatov, Aleksandr and Vanvlasselaer, Miguel",
    title = "{Bubble wall velocity: heavy physics effects}",
    eprint = "2010.02590",
    archivePrefix = "arXiv",
    primaryClass = "hep-ph",
    reportNumber = "SISSA 247/2020/FISI",
    doi = "10.1088/1475-7516/2021/01/058",
    journal = "JCAP",
    volume = "01",
    pages = "058",
    year = "2021"
}

@article{Baldes:2021vyz,
    author = "Baldes, Iason and Blasi, Simone and Mariotti, Alberto and Sevrin, Alexander and Turbang, Kevin",
    title = "{Baryogenesis via relativistic bubble expansion}",
    eprint = "2106.15602",
    archivePrefix = "arXiv",
    primaryClass = "hep-ph",
    reportNumber = "ULB-TH/21-09",
    doi = "10.1103/PhysRevD.104.115029",
    journal = "Phys. Rev. D",
    volume = "104",
    number = "11",
    pages = "115029",
    year = "2021"
}

@article{Zeldovich:1974uw,
    author = "Zeldovich, Ya. B. and Kobzarev, I. Yu. and Okun, L. B.",
    title = "{Cosmological Consequences of the Spontaneous Breakdown of Discrete Symmetry}",
    reportNumber = "SLAC-TRANS-0165, IPM-MOSCOW-15",
    journal = "Zh. Eksp. Teor. Fiz.",
    volume = "67",
    pages = "3--11",
    year = "1974"
}

@book{Vilenkin:2000jqa,
    author = "Vilenkin, A. and Shellard, E. P. S.",
    title = "{Cosmic Strings and Other Topological Defects}",
    isbn = "978-0-521-65476-0",
    publisher = "Cambridge University Press",
    month = "7",
    year = "2000"
}

@article{Abel:1995uc,
    author = "Abel, S. A. and White, P. L.",
    title = "{Baryogenesis from domain walls in the next-to-minimal supersymmetric standard model}",
    eprint = "hep-ph/9505241",
    archivePrefix = "arXiv",
    reportNumber = "RAL-TR-95-005, OUTP-95-17-P",
    doi = "10.1103/PhysRevD.52.4371",
    journal = "Phys. Rev. D",
    volume = "52",
    pages = "4371--4379",
    year = "1995"
}

@article{Planck:2015fie,
    author = "Ade, P. A. R. and others",
    collaboration = "Planck",
    title = "{Planck 2015 results. XIII. Cosmological parameters}",
    eprint = "1502.01589",
    archivePrefix = "arXiv",
    primaryClass = "astro-ph.CO",
    doi = "10.1051/0004-6361/201525830",
    journal = "Astron. Astrophys.",
    volume = "594",
    pages = "A13",
    year = "2016"
}

@inproceedings{Cline:2006ts,
    author = "Cline, James M.",
    title = "{Baryogenesis}",
    booktitle = "{Les Houches Summer School - Session 86: Particle Physics and Cosmology: The Fabric of Spacetime}",
    eprint = "hep-ph/0609145",
    archivePrefix = "arXiv",
    month = "9",
    year = "2006"
}

@inproceedings{Riotto:1998bt,
    author = "Riotto, Antonio",
    title = "{Theories of baryogenesis}",
    booktitle = "{ICTP Summer School in High-Energy Physics and Cosmology}",
    eprint = "hep-ph/9807454",
    archivePrefix = "arXiv",
    reportNumber = "CERN-TH-98-204",
    pages = "326--436",
    month = "7",
    year = "1998"
}

@article{Bodeker:2020ghk,
    author = "Bodeker, Dietrich and Buchmuller, Wilfried",
    title = "{Baryogenesis from the weak scale to the grand unification scale}",
    eprint = "2009.07294",
    archivePrefix = "arXiv",
    primaryClass = "hep-ph",
    reportNumber = "DESY 20-141, DESY-20-141",
    doi = "10.1103/RevModPhys.93.035004",
    journal = "Rev. Mod. Phys.",
    volume = "93",
    number = "3",
    pages = "035004",
    year = "2021"
}

@article{Planck:2018vyg,
    author = "Aghanim, N. and others",
    collaboration = "Planck",
    title = "{Planck 2018 results. VI. Cosmological parameters}",
    eprint = "1807.06209",
    archivePrefix = "arXiv",
    primaryClass = "astro-ph.CO",
    doi = "10.1051/0004-6361/201833910",
    journal = "Astron. Astrophys.",
    volume = "641",
    pages = "A6",
    year = "2020",
    note = "[Erratum: Astron.Astrophys. 652, C4 (2021)]"
}

@article{Yeh:2022heq,
    author = "Yeh, Tsung-Han and Shelton, Jessie and Olive, Keith A. and Fields, Brian D.",
    title = "{Probing physics beyond the standard model: limits from BBN and the CMB independently and combined}",
    eprint = "2207.13133",
    archivePrefix = "arXiv",
    primaryClass = "astro-ph.CO",
    reportNumber = "UMN-TH-4125/22, FTPI-MINN-22/16",
    doi = "10.1088/1475-7516/2022/10/046",
    journal = "JCAP",
    volume = "10",
    pages = "046",
    year = "2022"
}

@article{Burns:2023sgx,
    author = "Burns, Anne-Katherine and Tait, Tim M. P. and Valli, Mauro",
    title = "{PRyMordial: the first three minutes, within and beyond the standard model}",
    eprint = "2307.07061",
    archivePrefix = "arXiv",
    primaryClass = "hep-ph",
    reportNumber = "UCI-HEP-TR-2023-07, YITP-SB-2023-16",
    doi = "10.1140/epjc/s10052-024-12442-0",
    journal = "Eur. Phys. J. C",
    volume = "84",
    number = "1",
    pages = "86",
    year = "2024"
}

@article{Pitrou:2019nub,
    author = "Pitrou, Cyril and Coc, Alain and Uzan, Jean-Philippe and Vangioni, Elisabeth",
    editor = "Kawabata, T. and others",
    title = "{Precision Big Bang Nucleosynthesis with the New Code PRIMAT}",
    eprint = "1909.12046",
    archivePrefix = "arXiv",
    primaryClass = "astro-ph.CO",
    doi = "10.7566/JPSCP.31.011034",
    journal = "JPS Conf. Proc.",
    volume = "31",
    pages = "011034",
    year = "2020"
}

@article{Fromme:2006wx,
    author = "Fromme, Lars and Huber, Stephan J.",
    title = "{Top transport in electroweak baryogenesis}",
    eprint = "hep-ph/0604159",
    archivePrefix = "arXiv",
    reportNumber = "CERN-PH-TH-2006-064, BI-TP-2006-10",
    doi = "10.1088/1126-6708/2007/03/049",
    journal = "JHEP",
    volume = "03",
    pages = "049",
    year = "2007"
}

\end{document}